\documentclass[a4paper,english,aps,pre,reprint,showpacs,preprintnumbers]{revtex4-1}
\usepackage[latin9]{inputenc}
\usepackage{color}
\usepackage{babel}
\usepackage{amsmath}
\usepackage{amssymb}
\usepackage{graphicx}
\usepackage[unicode=true,
 bookmarks=false,
 breaklinks=false,pdfborder={0 0 1},backref=section,colorlinks=true]
 {hyperref}
\hypersetup{
 citecolor=blue}

\graphicspath{{./}{./FIGS/}}


\begin{document}

    \title{Dispersive instabilities in Passively Mode-Locked Integrated External-Cavity Surface-Emitting Lasers}

    \author{Christian Schelte$^{1,2}$}
    \author{Denis Hessel$^{1,2}$}
    \author{Julien Javaloyes$^{1}$}
    \author{Svetlana V. Gurevich$^{1,2,3}$}
    \email{gurevics@uni-muenster.de}
    \affiliation{$^{1}$ Departament de F\'isica, Universitat de les Illes Balears \& Institute of Applied Computing and Community Code (IAC-3), Cra.\,\,de Valldemossa, km 7.5, E-07122 Palma de Mallorca, Spain}
    \affiliation{$^{2}$Institute for Theoretical Physics, University of M\"unster, Wilhelm-Klemm-Str. 9, D-48149 M\"unster, Germany}
    \affiliation{$^{3}$Center for Nonlinear Science (CeNoS), University of M\"unster, Corrensstrasse 2, D-48149 M\"unster, Germany}
    \date{\today}
    
\selectlanguage{english}%
\begin{abstract}
We analyze the dynamics of passively mode-locked integrated external-cavity
surface-emitting Lasers (MIXSELs) using a first-principle dynamical
model based upon delay algebraic equations. We show that the third
order dispersion stemming from the lasing micro-cavity induces a train
of decaying satellites on the leading edge of the pulse. Due to the
nonlinear interaction with carriers, these satellites may get amplified
thereby destabilizing the mode-locked states. In the long cavity regime,
the localized structures that exist below the lasing threshold are
found to be deeply affected by this instability. As it originates
from a global bifurcation of the saddle-node infinite period type,
we explain why the pulses exhibit behaviors characteristic of excitable
systems. Using the multiple time-scale and the functional mapping
methods, we derive rigorously a master equation for MIXSELs in which
third order dispersion is an essential ingredient. We compare the
bifurcation diagram of both models and assess their good agreement. 

\end{abstract}
\maketitle

\section{Introduction}

In spite of its early discovery in 1965 \citep{MC-APL-65}, the passive
mode-locking (PML) of lasers is still a subject of intense research,
not only due to its important technological applications \citep{lorenser04,keller96},
but also because it involves the self-organization of a large number
of laser modes that experience an out-of-equilibrium phase transition
\citep{GP-PRL-02,WRG-PRL-05}.

Vertical external-cavity surface-emitting semiconductor lasers (VECSELs)
are prominent laser sources in various industrial and scientific applications
that require high output power and good beam quality \citep{Innoptics,DMB-JSTQE-13,lidar,lidar2}.
Here, the PML phenomenon is obtained by closing the external cavity
with a semiconductor saturable absorber mirror (SESAM); the intensity-dependent
losses promote pulsed over continuous wave emission and lead to pulses
that are typically in the picosecond range, see \citep{haus00rev,AJ-BOOK-17}
for reviews.

Among the wealth of dynamical regimes encountered in such complex
photonic systems involving coupled cavities, the regular pulsating
regimes have direct applications, e.g., for compact spectroscopy \citep{LMW-SCI-17}
and metrology \citep{UHH-NAT-02}. Designs based upon the VCSEL-SESAM
geometry yield output powers that evolved from $200\,$mW \citep{hoogland00,haring01,haring02}
towards $20\,$W \citep{RRH-OL-08} and, recently, transform limited
pulses in the 100~fs range with peak power of 500~W \citep{WLM-OptA-16}
were obtained with repetition rate in the GHz range.

A new type of nonlinear cavity appeared in the last decade, the so-called
Mode-Locked Integrated External-Cavity Surface-Emitting Lasers (MIXSEL)
\citep{MBR-APB-07,RWM-OE-10} in which both the gain and the saturable
absorber share the same micro-cavity. Coupling the MIXSELs to an external
mirror provides for the essential optical feedback that defines the
repetition rate of the pulse train. 

More generally, the mode-locking of VECSELs 
is based upon micro-cavities operated in reflection: one or two for 
MIXSELs and VCSEL-SESAMs, respectively. 
The rich PML dynamics and the multi-pulses regimes can be controlled
e.g., with time-delayed optical feedback \citep{JNS-PRE-16,JKL-CHA-17},
coherent optical injection \citep{AHP-JOSAB-16} or photonic crystal
structures \citep{SCB-PRL-19}. In addition, carrier dynamics
in multi-level active materials such as quantum dots \citep{RBM-JQE-11,BSR-BOOK-14}
or sub-monolayer quantum dots \citep{AWN-PRAp-18} leads to rich
behaviors. The cavity geometry also proved its relevance and peculiar
pulse clusters appear in V-shaped cavities \citep{MLA-OE-18,HML-PRAp-19}.

While it was commonly accepted that semiconductor mode-locked lasers
could not emit pulse trains at rates well below the GHz, a regime of
temporal localization allowing arbitrary low repetition rates and
individual pulse addressing was disclosed \citep{MJB-PRL-14,CJM-PRA-16,JCM-PRL-16,CSV-OL-18,YRS-PRL-19}.
This transition from PML towards addressable temporal localized structures
(TLSs) could have applications for dense frequency
combs generation \citep{SPH-NAP-12} and all optical data processing \citep{K-NAT-03}.
To add to its technological relevance, the long cavity regime, in
which the photon round-trip is longer than the semiconductor gain
recovery time, is compatible with spatial confinement; stable three-dimensional
light bullets were predicted in broad area micro-cavities \citep{J-PRL-16,GJ-PRA-17,SJG-OL-18}.

Recently, a new kind of instability for the pulse trains obtained
in the long cavity regime was experimentally observed in mode-locked
VECSELs \citep{SCM_PRL_19}. At its core, these unstable pulsating
regimes results from the face-to-face coupling of the micro-cavities
containing the gain and the saturable absorber media. Operated in
reflection, the gain and absorber micro-cavities behave as dispersive
Gires-Tournois interferometers (GTI) \citep{GT-CRA-64}. After several
round-trips, the third order dispersion (TOD) induced by the micro-cavities
give rise to serrated wave forms that consists in a decaying sequence
of satellites accumulating in front of the leading edge of the pulse.
It was shown that these satellites may become unstable leading to
a low frequency modulation of the pulse envelope on a slow time scale,
typically on the order of hundreds of round-trips. Understanding this
regime is of paramount importance in order to avoid time and amplitude
jitter and spectral broadening of the lines forming the frequency
comb of the PML regimes.

In this manuscript, we predict that such dispersive instabilities
may also occurs in MIXSELs. While MIXSELs micro-cavities
contain both the gain and the saturable absorption, they are also
operated in reflection and it stands to reason that they shall exhibit
the same dispersive behavior characteristics of GTIs. Finally, due
to its simpler geometry, the MIXSEL proves to be a particularly instructive
minimal model for our analysis.

We present in Sec.\ref{sec:modeleq} a first principle model for the
MIXSEL. Following the method of \citep{MB-JQE-05}, it consists in
solving analytically the field propagation in the linear sections
of the micro-cavity, while considering the gain and absorber sections
as nonlinear boundary conditions. The boundary conditions for the
field at the micro-cavity/air interface together with the optical
feedback from the external mirror impose the structure of the model
as time-delayed algebraic equations (DAEs). We propose
to conduct what we believe to be the first bifurcation analysis of
such DAEs model in Photonics. We demonstrate in Sec. \ref{sec:sat_instab}, how
a series of decaying satellites on the leading edge of the pulse causes
an instability of the pulse train and how a global bifurcation with
features of excitability as well as more intricate oscillating dynamics
can appear in Sec. \ref{sec:excitable} and \ref{sec:mixed}, respectively.
Using the rigorous formalism of the multiple time-scale \citep{NF-BOOK}
as well as the newly developed functional mapping method \citep{SJG-OL-18},
we complement our analysis in Sec.\ref{sec:pde} with the derivation
of a master equation governing the slow pulse evolution over the round-trip
time scale, similar to the so-called Haus master equation \citep{haus00rev},
and that takes the form of a generalized complex Ginzburg-Landau partial
differential equation (PDE). The latter allows to identify why TOD
has such an important effect in the dynamics of MIXSELs. Finally,
using modern continuation techniques \citep{uecker2014}, we compare
the results of the PDE with that of the DAE model and find in which
conditions a good agreement can be obtained.


\section{Model system}

\label{sec:modeleq}

\begin{figure}
\includegraphics[width=0.4\textwidth]{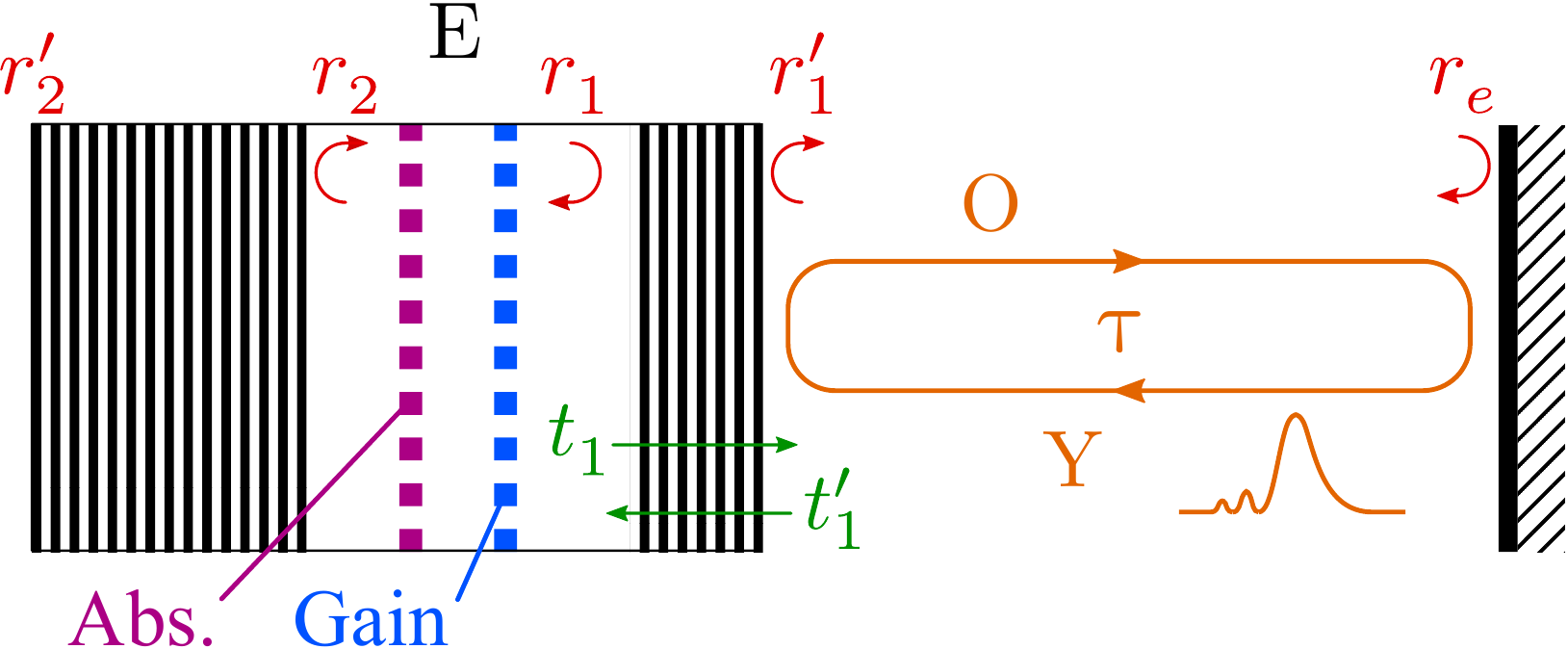} \caption{(color online) A schematic of the MIXSEL configuration, where both
gain (blue) and saturable absorption (magenta) are contained in the
same microcavity. $E$ denotes the field amplitude in the active region.
The output and injection fields in the external cavity are represented
by $O$ and $Y$, respectively. The external cavity round-trip time
is $\tau$.}
\label{fig:setup} 
\end{figure}

The schematic setup of a MIXSEL system is depicted in Fig.~\ref{fig:setup}.
The gain and the absorber media are enclosed into micro-cavities whose
length is of the order of the lasing wavelength. The two mirrors of
the micro-cavity provide additional degrees of freedom for controlling
the light-matter interaction. The interaction strength with the active
medium ---that is only a few tens of nanometers long--- can be dramatically
increased, as the expense of the available bandwidth, by using high-Q
cavities. Similarly, the effective saturation of the active material
can be increased (or decreased) by using resonant or anti-resonant
cavity designs, respectively. Because of the vast scale separation
between the external cavity length and that of the micro-cavity, the
natural framework for our analysis is that of time-delayed systems.
The latter appear not only as natural modeling approaches for PML
\citep{MB-JQE-05,VT-PRA-05} but in many branches of physics. Delayed
systems have strong links with spatially extended systems such as
the Ginzburg-Landau equation \citep{GP-PRL-96,K-CMMP-98} and lead to 
rich dynamical behaviors \citep{GMZ-PRE-13,LPM-PRL-13,MGB-PRL-14,YG-PRL-14,EJW-CHA-17},
see \citep{YG-JPA-17} for a review.

We consider the case of a resonant cavity and we denote by $E$ the
micro-cavity field over an antinode and $Y$ the field in the external
cavity. The output field is denoted by $O$, $\tau$ is the external
cavity round trip time, whereas $r_{1,2}$ are the top and bottom
Distributed Bragg Reflector (DBR) reflectivities, $t_{1}$ is the
transmission coefficient of the top DBR and $r_{e}$ is the external
mirror reflectivity. We follow the approach of \citep{MB-JQE-05}
that consists in solving the field propagation in the linear sections
of the micro-cavity. That way one obtains a dynamical model linking
the two fields $E$ and $Y$. Their coupling is achieved considering
the transmission and reflection coefficients of the top DBR. After
normalization, one obtains the rate equations for the field $E$,
the gain $N_{1}$ and absorber $N_{2}$ population inversions as 
\begin{align}
\dot{E} & =\left[\left(1-i\alpha_{1}\right)N_{1}+\left(1-i\alpha_{2}\right)N_{2}-1\right]E+hY,\label{eq:DAE1}\\
\dot{N}_{1} & =\gamma_{1}\left(J_{1}-N_{1}\right)-\left|E\right|^{2}N_{1},\label{eq:DAE2}\\
\dot{N}_{2} & =\gamma_{2}\left(J_{2}-N_{2}\right)-s\left|E\right|^{2}N_{2},\label{eq:DAE3}\\
Y & =O\left(t-\tau\right)=\eta\left[E\left(t-\tau\right)-Y\left(t-\tau\right)\right].\label{eq:DAE4}
\end{align}
We scaled Eqs.~(\ref{eq:DAE1}-\ref{eq:DAE4}) by the photon lifetime
in the micro-cavity $\tau_p$, and $\alpha_{1}$ and $\alpha_{2}$
are the linewidth enhancement factors of the gain and absorption,
respectively. We set the bias and the recovery time in the gain as
$\left(J_{1},\gamma_{1}\right)$ and in the absorber section as $\left(J_{2},\gamma_{2}\right)$,
respectively. The ratio of the gain and absorber saturation intensities
is $s$.

The cavity enhancement due to the high reflectivity mirrors can be
scaled out, making that $E$ and $Y$ are of the same order of magnitude.
This scaling has the additional advantage of simplifying the input-output
relation of the micro-cavity; using Stokes relations, we find that
it reads $O=E-Y$. The minus sign represents the $\pi$ phase shift
of the incoming field $Y$ upon reflection from the top DBR. After
a round-trip in the external cavity of duration $\tau$, the output
field $O\left(t-\tau\right)$ is re-injected with an attenuation factor
$\eta=r_{e}\exp\left(\omega_{0}\tau\right)$, with $\omega_{0}\tau$
the propagation phase, defining $\omega_{0}$ as the carrier frequency
of the field. The coupling between $E$ and $Y$ is given in Eq.~(\ref{eq:DAE4})
by a delayed algebraic equation (DAE), that takes into account the
multiple reflections in the external cavity. In the limit of a very
low external mirror reflectivity $\eta\ll1$, one would truncate the
infinite hierarchy generated by Eq.~(\ref{eq:DAE4}) to obtain $Y=\eta E\left(t-\tau\right)+\mathcal{O}\left(\eta^{2}\right)$
leading to the so-called Lang-Kobayashi model \citep{LK-JQE-80}.
Yet, for mode-locked configurations $\eta=\mathcal{O}\left(1\right)$
and the multiple reflections in the external cavity must be taken
into account. Instead of considering an infinite number of delayed
terms in Eq.~(\ref{eq:DAE1}) with values $\tau,\,2\tau,\,\cdots,\,n\tau$,
the DAE given by Eq.~(\ref{eq:DAE4}) allows for an elegant representation
of the strongly coupled cavity dynamics without needing an \emph{a
priori} truncation.

The coupling efficiency of the external field $Y$ into the micro-cavity
is given by the parameter $h=\left(1+\left|r_{2}\right|\right)\left(1-\left|r_{1}\right|\right)/\left(1-\left|r_{1}r_{2}\right|\right)$.
There exist three instructive limit cases for the coupling parameter
$h$ that correspond to certain types of devices: A non-transmitting
top DBR $\left|r_{1}\right|=1$ yields $h=0$, making the cavity
equivalent to a perfect (linear) mirror. Equal reflectivities for
both DBRs $\left|r_{1}\right|=\left|r_{2}\right|$ yield $h=1$ and
correspond to a symmetric Fabry-Perot cavity. Finally, a fully reflecting
bottom DBR $\left|r_{2}\right|=1$ yields $h=2$, which corresponds
to the GTI case \citep{GT-CRA-64}. Gires-Tournois interferometers
are known for inducing controllable second order dispersion and they
are used as optical pulse shaping elements. Resonant photons transmitted
into the micro-cavity will remain on average for the photon lifetime.
When transmitted back into the external cavity they will have collected
a phase difference with respect to the off-resonance photons that
are directly reflected upon the top DBR. Note that this phase shift
is a function of the detuning of the photons with respect to the closest
micro-cavity mode. The recombination of various wavelength in the
external cavity leads to dispersion. This process is fully captured
by Eq.~(\ref{eq:DAE1}-\ref{eq:DAE4}). Second order dispersion is
typically the dominating effect and its amount is tunable by choosing
the detuning. Using red or blue detuning one can achieve either normal
or anomalous dispersion while around resonance TOD becomes the leading
term as the second order contribution vanishes and switches sign.
Gires-Tournois interferometers are designed to conserve the photon
number using high reflective bottom mirrors and therefore yield purely
dispersive spectrum in models such as given by Eqs.~(\ref{eq:DAE1}-\ref{eq:DAE4}),
see \citep{SCM_PRL_19} for more details.

In order to achieve directional emission and low losses the bottom
DBRs of VCSELSs are optimized towards $\left|r_{2}\right|\rightarrow1$,
i.e., they are well approximated by the GTI regime and $h\rightarrow2$.
We set the photon lifetime as $\tau_p=3\,$ps which corresponds
to a Full Width at Half Maximum (FWHM) of $(\pi\tau_p)^{-1}=106\,$GHz.
The gain and absorber lifetimes are $1\,$ns and $30\,$ps, respectively,
while we set the round-trip time in the cavity to $3\,$ns, hence
$\left(\gamma_{1},\gamma_{2},\tau\right)=\left(0.003,0.1,1000\right).$
If not stated otherwise, the other parameters are $(J_{2},\eta,s,h)=(-0.5,0.7,10,2)$.


\section{Satellite instability}

\label{sec:sat_instab}

Due to TOD from the GTI-like micro-cavity pulses can have a series
of decaying satellites on the leading edge that was found to cause
an instability of the pulse train \citep{SCM_PRL_19}. Without linewidth
enhancement factors $\alpha_{1}=\alpha_{2}=0$ the satellites are
most clearly developed because of the absence of chirp and their instability
can be better understood starting from this situation. In a real semiconductor
medium the change in carrier density along the pulse profile causes
a varying detuning with respect to the micro-cavity resonance due
to the alpha factors. The resulting mixture between chirp and dispersion
creates a more involved dynamics that will be discussed later in this
manuscript.

    \begin{figure}
    \includegraphics[viewport=1bp 0bp 3918bp 3623bp,clip,width=0.5\textwidth]{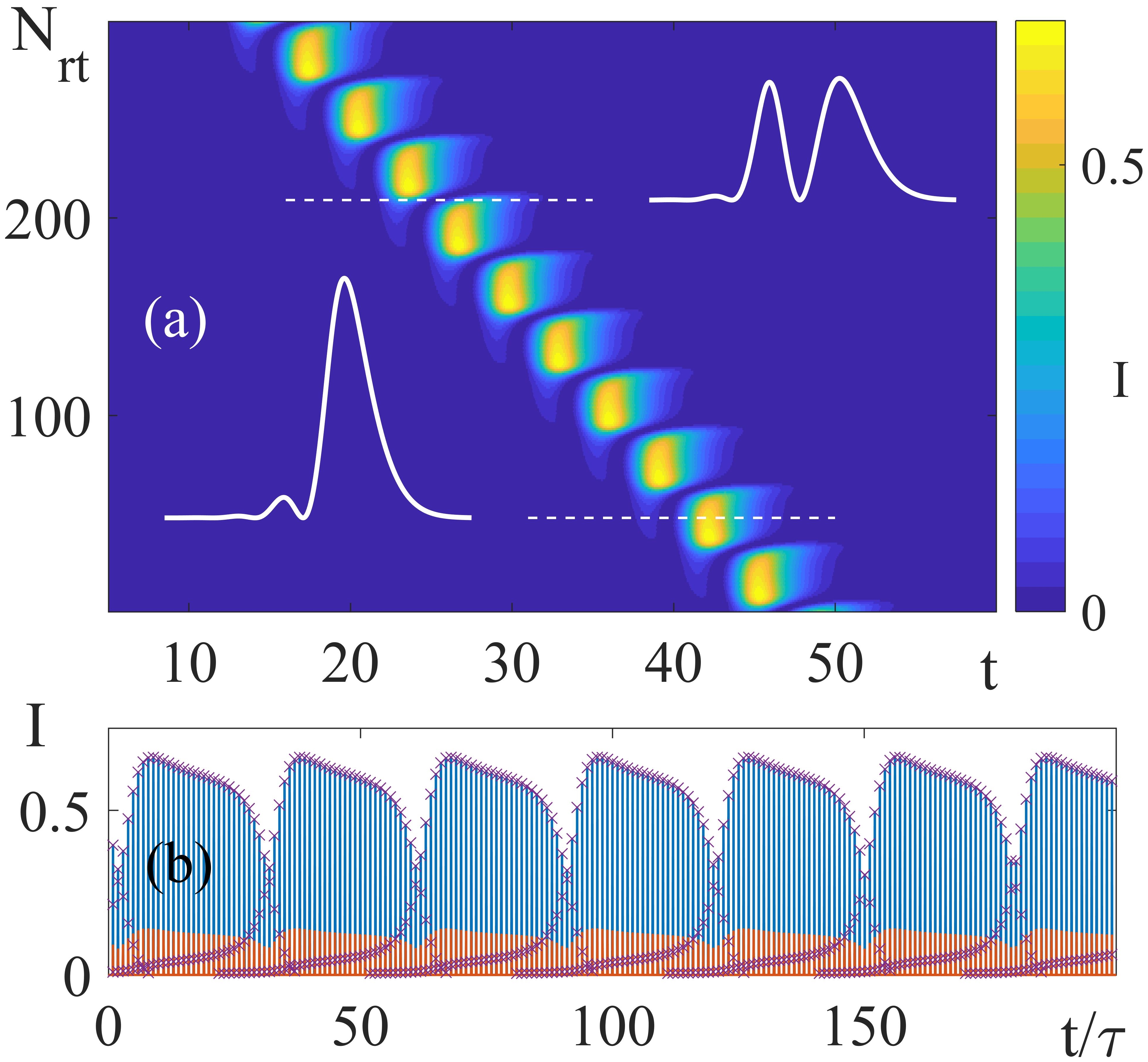}
        \caption{(color online)
            Pseudo-space-time diagram (a) and time trace (b) for the pulse train in the unstable satellite regime obtained from DNSs of Eqs.~(\ref{eq:DAE1}-\ref{eq:DAE4}). The pulse intensity for $E$ (blue) and $Y$ (orange) fields is shown.  The purple crosses at the intensity peaks illustrate the creation-annihilation cycle.  For sufficiently large gain the largest satellite is amplified, eventually replacing its parent pulse.  Parameters are $(J_1,\,\alpha_1,\,\alpha_2)=(0.65,\,0,\,0).$
        }
        \label{fig:clean_sat}
    \end{figure}

We operated in the regime of localization where the pulses are temporal
localized states that appear below the lasing threshold bias
defined as $J_{1}^{\mathrm{th}}$ . As detailed in \citep{MJB-PRL-14},
the TLSs appear via a saddle-node bifurcation of limit cycles. Sufficiently
close to the lasing threshold, the main pulses and therefore their
parasitic satellites become large enough to bleach the absorber and
open the net gain window prematurely. As a consequence, they grow
exponentially from one round-trip towards the next while the parent
pulse meets an increasingly depleted gain carrier density and eventually
dies out. It is replaced by its satellite in front, resulting in forward
leaping motion that can best be seen in a pseudo-space-time representation
which is shown in Fig.~\ref{fig:clean_sat}(a) where $N_{rt}$ is
the number of round-trips. The corresponding temporal trace obtained
from direct numerical simulations (DNSs) of Eqs.~(\ref{eq:DAE1}-\ref{eq:DAE4})
is depicted in Fig.~\ref{fig:clean_sat}(b). It demonstrates how
this cycle of creation and annihilation leads to a low frequency modulation
of the pulse train (see the purple crosses).

    \begin{figure}
\includegraphics[viewport=90bp 30bp 820bp 360bp,clip,width=0.48\textwidth]{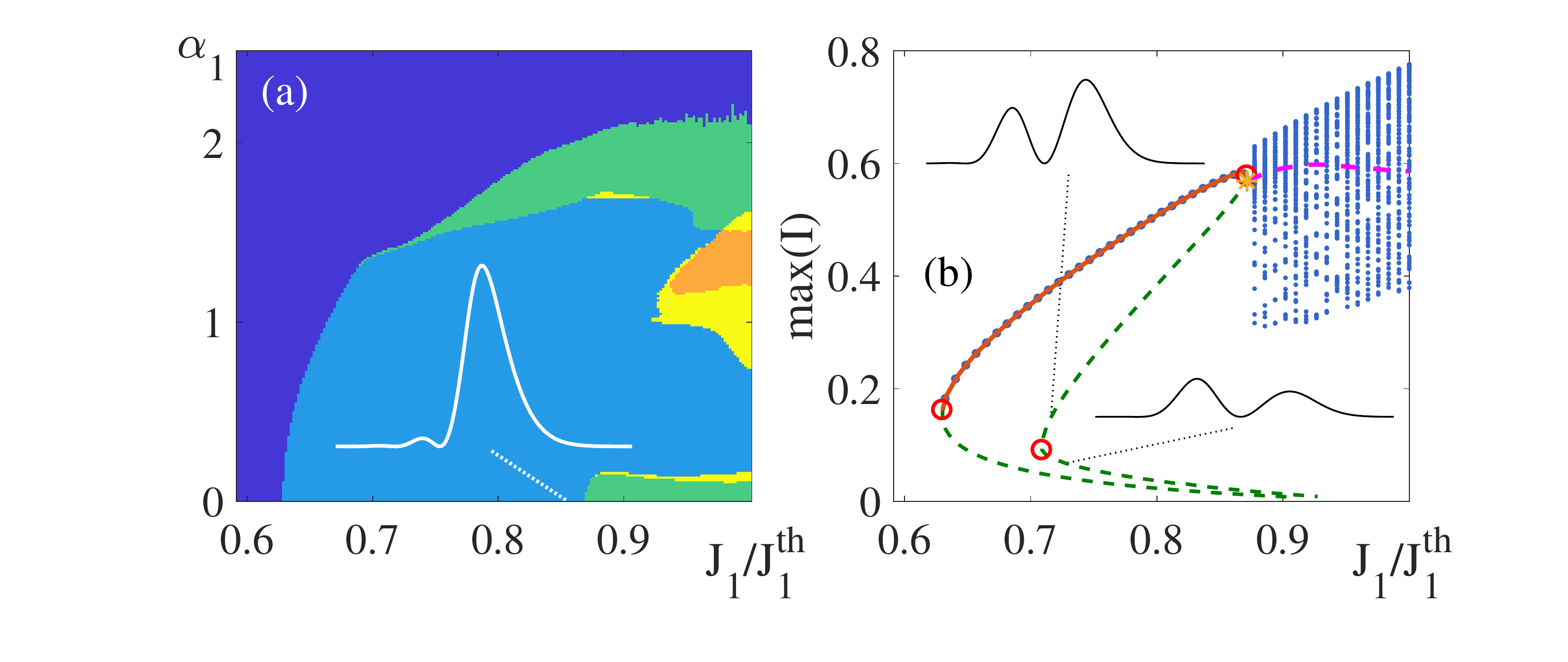}
        \caption{(color online)
            (a) Bifurcation diagram in the $(J_1,\alpha_1)$-plane obtained with DNSs of Eqs.~(\ref{eq:DAE1}-\ref{eq:DAE4}) for $\alpha_2=0$.  The off, stable pulsating, and quasi-periodic pulsating regimes are shown in dark blue, light blue and green, respectively.  Bistable regions are orange and yellow for stable and oscillating solutions.  Pure satellite instability is observed for low $\alpha_1$ whereas trailing edge AH instability for high $\alpha_1$ values.  Both can mix forming the region around $\alpha_1\approx1$ close to threshold that is bistable with stable pulses. (b) Branch of a single pulse solution showing $\mathrm{max(}I)$ as a function of the scaled gain bias $J_1$, superposing results from DNS (blue points) and path-continuation for $\alpha_1=0$.  Orange solid line shows the stable part of the branch which becomes unstable (green dashed line) in a fold of the limit cycle (red circles), i.e., the satellite instability does not a stem from secondary AH bifurcation.  After the fold on the unstable branch there is a supercritical pitchfork bifurcation (yellow star) that gives birth to another unstable branch (dashed magenta).
        }
        \label{fig:bif_diag}
    \end{figure}

A full bifurcation analysis of the DAE system (\ref{eq:DAE1}-\ref{eq:DAE4})
is out of the scope of the current manuscript. The main reason being
that the stability analysis of periodic solutions of DAEs is not
possible at the moment, even using the most advanced continuation
softwares such as DDE-BIFTOOL \citep{DDEBT}. Yet, we show in Fig.~\ref{fig:bif_diag}(a)
a numerical two parameter bifurcation diagram in the $\left(J_{1},\alpha_{1}\right)$
plane for the single pulse solution. The stable region and off solution
are depicted in light and dark blue, respectively, in Fig.~\ref{fig:bif_diag}(a).
For high values of the linewidth enhancement factor of the gain $\alpha_{1}$,
a quasi-periodic instability due to self-phase modulation is found,
similar to that discussed in \citep{SJG-PRA-18}. This instability
was found to be a local secondary Andronov-Hopf (AH) bifurcation.
In addition to these regimes, that exists in the generic mode-locked
ring laser model of \citep{VT-PRA-05}, a new instability induced
by TOD is found for low values of $\alpha_{1}$. It corresponds to
the regime depicted in Fig.~\ref{fig:clean_sat}. Both unstable regions
are depicted in green. In addition, around $\alpha_{1}\approx1$,
there exists a bistable region close to threshold between a stable PML 
regime with higher intensity pulses and low frequency modulated PML; this latter quasi-periodic
dynamics has the characteristics of both instabilities found for low
and large $\alpha_{1}$; it is discussed in more detail in Section~\ref{sec:mixed}.
The bistable region is depicted in orange where stable and modulated
pulses are stable and yellow where only one solution is stable. Note
that parts along the borders of the pure instabilities are also bistable
with a regular PML solution with smaller pulses.

We show in Fig.~\ref{fig:bif_diag}(b) the branch for the single
TLS solution obtained using DDE-BIFTOOL.
At $\alpha_{1}=0$, the solution branch folds three times with the
second fold at a critical value of the current $J_{1}^{c}\sim0.86J_{1}^{\mathrm{th}}$
that coincides with the onset of the satellite instability. The results
from the corresponding DNSs are superposed (blue points), indicating
the all the values of the pulse maxima integrated over many round-trips.
Shortly after the second fold, a second unstable branch (dashed magenta)
appears in a pitchfork bifurcation. Above $J_{1}^{c}$, no stable
single pulse solution branch exists and one observes the low frequency
periodic dynamics discussed in Fig.~\ref{fig:clean_sat}.

    \begin{figure}
    \includegraphics[viewport=0bp 0bp 290bp 165bp,clip,width=0.45\textwidth]{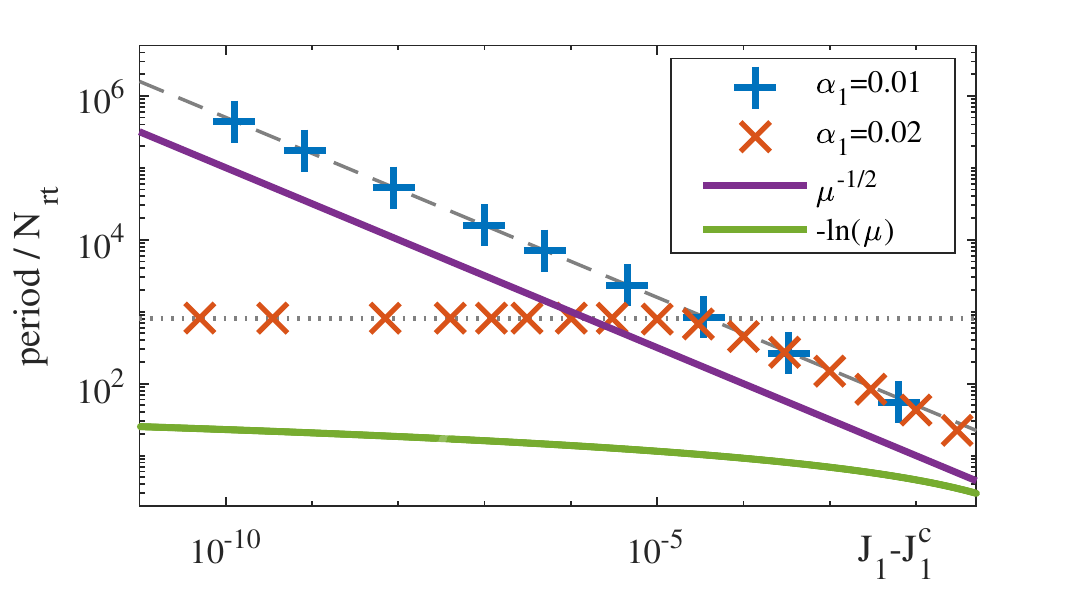}
        \caption{(color online)
            The dependence of the period length on the distance to the bifurcation point $J_1^c$.  The crosses denote the period obtained from DNSs of Eqs. (\ref{eq:DAE1}-\ref{eq:DAE4}) for $\alpha_1=0.01$ (blue) and $\alpha_1=0.02$ (orange).  The straight lines correspond to the theoretically predicted scaling behavior for a homoclinic (green) as well as for a SNIPER (purple) bifurcation.
        }
        \label{fig:scaling}
    \end{figure}

It has to be noted that for $\alpha_{1}\ll1$ this instability does
not stem from a local Andronov-Hopf bifurcation but from a global
bifurcation. The limit cycle is born with infinite period at the second
fold of the TLS branch in Fig.~\ref{fig:bif_diag}(b). The period
scaling is presented in Fig.~\ref{fig:scaling} where we show the
period evolution for two exemplary values of $\alpha_{1}$ (blue and
orange crosses) and the characteristic scalings for saddle-node infinite
period ($\mu^{-1/2}$) and homoclinic ($\ln\mu^{-1}$) bifurcations
as a function of the distance to the bifurcation point $J_{1}^{c}$.
The results reveal that for small values of $\alpha_{1}$ the satellite
instability can be identified to be of the saddle-node infinite period
(SNIPER) type. However, for increasing $\alpha_{1}$ the SNIPER bifurcation
changes into a local AH bifurcation (cf. the scaling for $\alpha_{1}=0.02$).

\section{Excitability}

\label{sec:excitable}

    \begin{figure}
\includegraphics[viewport=10bp 0bp 270bp 225bp,clip,width=0.5\textwidth]{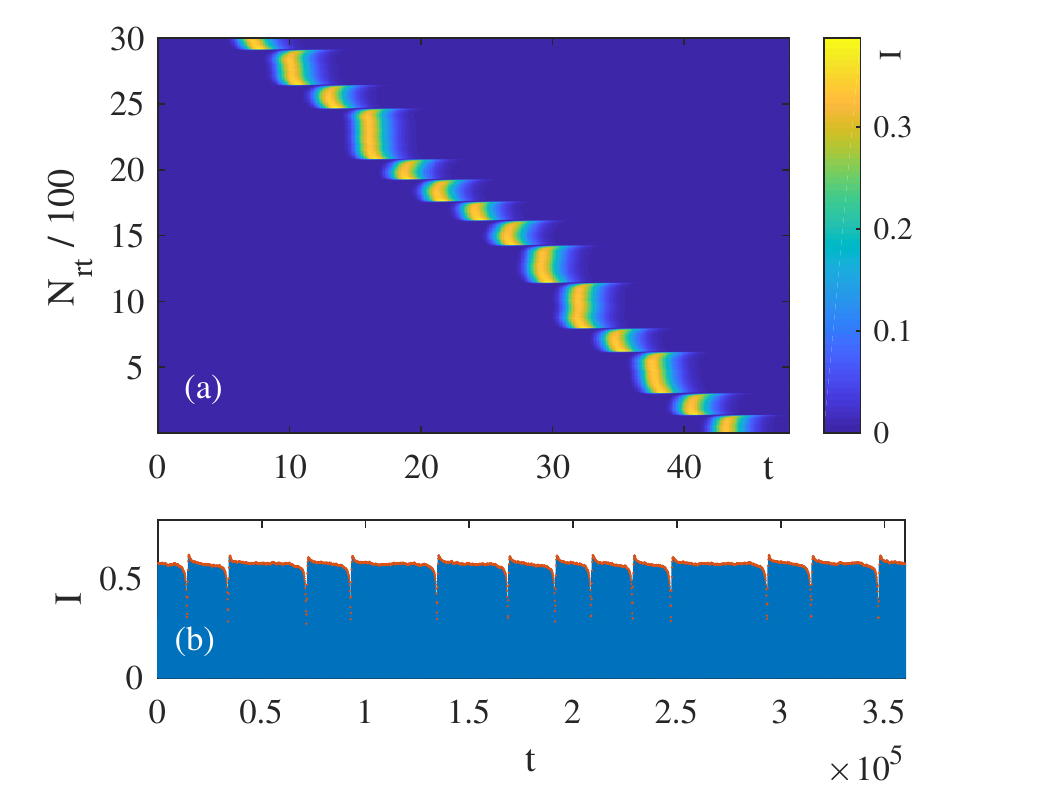}
        \caption{(color online)
            (a) Pseudo-space-time diagram for a single pulse obtained in DNSs of (\ref{eq:DAE1}-\ref{eq:DAE4}) showing excitable behavior.  Close to the SNIPER bifurcation point, satellites grow slowly as they are barely able to bleach the absorber sufficiently.  High noise has a very strong influence on the blow-up time in this situation.  (b) Corresponding time trace of the pulse intensity with the pulse maxima highlighted in orange.
        }
        \label{fig:excitability}
    \end{figure}
    
Under the influence of noise the satellite instability exhibits dynamical
behavior characteristics of excitable systems. To trigger the replacement
of the parent pulse a satellite must have sufficient intensity to
bleach the absorber and open an early net-gain window. Close, yet
below, to the critical energy, noise can help or hinder the satellite
emergence by adding or subtracting some energy, respectively. As a
result, the period of the pulse eruptions can become highly irregular
and we present in Fig.~\ref{fig:excitability} an exemplary pseudo-space-time
diagram for a single pulse (a) and the corresponding time trace (b)
in the excitable regime. By all accounts, this is certainly in its
irregular form that such a regime is most likely going to appear in
an experimental situation, where noise is unavoidable.


    \begin{figure}
        \includegraphics[width=.48\textwidth]{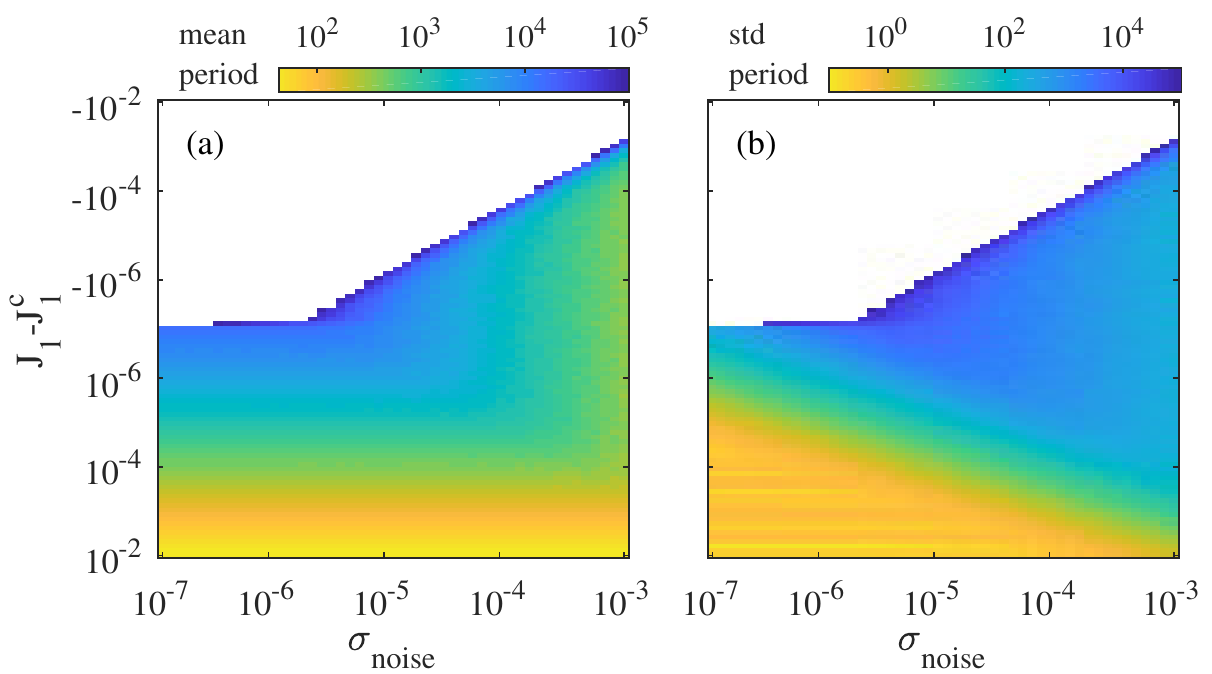}
        \caption{(color online)
            The dependence of the mean value (a) and standard deviation (b) of the period of the satellite instability as a function of the noise parameter $\sigma_\mathrm{noise}$ and the distance of $J_1$ to the bifurcation point $J_1^{c}$. Parameters are $(\alpha_1,\alpha_2)=(0,0)$.
        }
        \label{fig:statistics}
    \end{figure}

In Fig.~\ref{fig:statistics} we show the statistical properties
of the satellite instability under the influence of gaussian noise
with a standard deviation of $\sigma_{\mathrm{noise}}$ for varying
distances of the gain bias $J_{1}$ from the critical value $J_{1}^{c}$
corresponding to the onset of the SNIPER bifurcation. Panel (a) shows
the mean value and panel (b) the standard deviation of the period
of the satellite eruptions. At each point of the diagrams hundred
of consecutive eruptions were analyzed. To make such calculations
feasible, the simulations were performed in the long delay limit using
the functional mapping approach \citep{SJG-OL-18}. One can see how
noise enables eruptions below the critical pump rate. The stronger
the noise the less gain bias is necessary for this. The cutoff visible
in the data is arbitrary, as the necessary integration times grows
very fast and becomes a limiting factor to the feasibility of DNSs.
Statistically, eruptions can happen anywhere, albeit very rarely depending
on the parameters. For increasingly strong noise the observed average
period decreases and depends little on the actual gain bias. Similarly,
for strong pumping the noise hardly changes the mean period. The standard
deviation of the eruptions is very large at the onset below critical
pumping due to noise. Here the eruptions are rare and the excitability
is most visible. Otherwise the variance decreases with both increasing
gain bias and decreasing noise level.

\section{Combined instabilities}

\label{sec:mixed}

\begin{figure}
\includegraphics[width=0.49\textwidth]{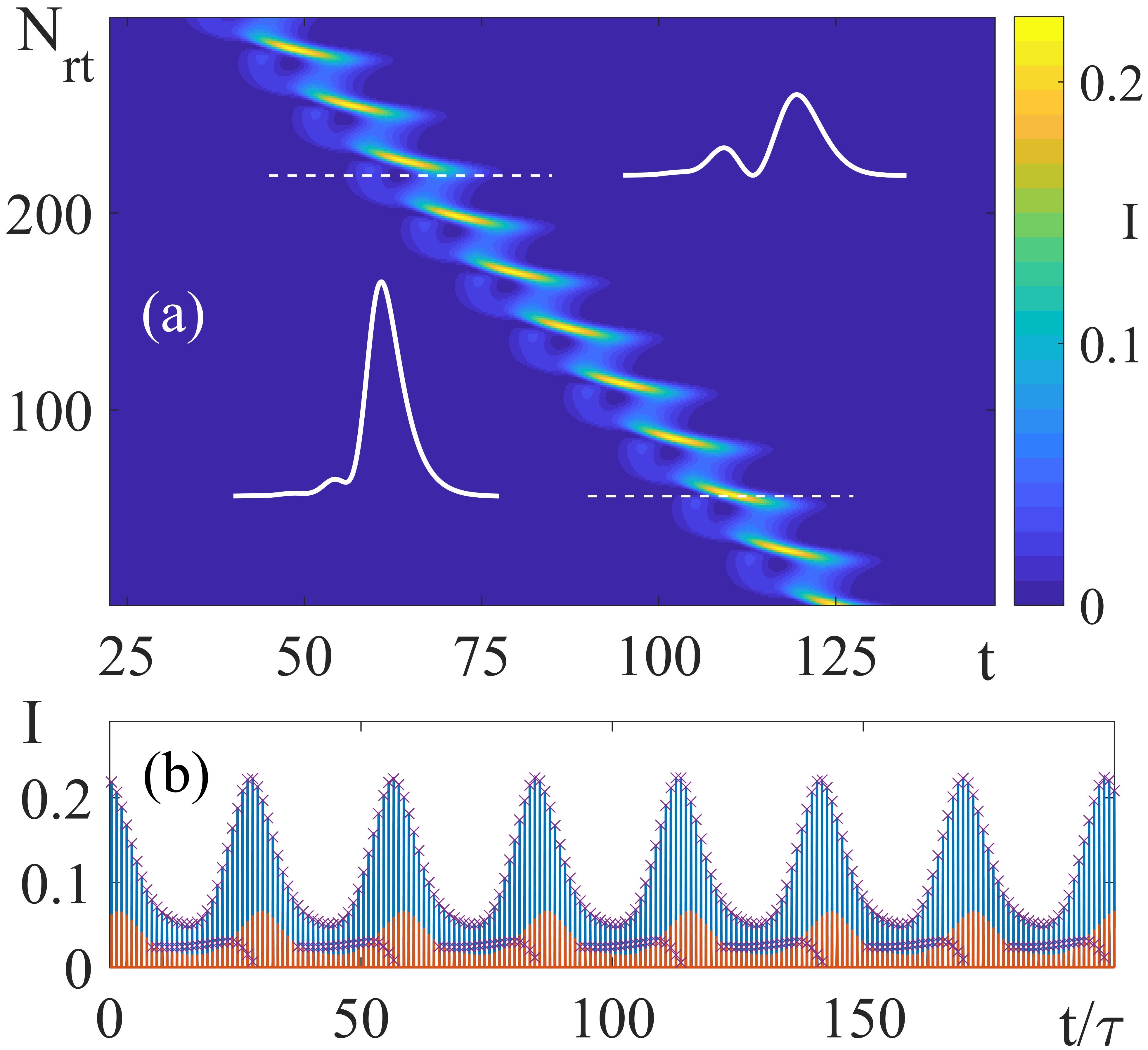} 
\caption{(color online) Pseudo-space-time diagram (a) and corresponding time
trace (b) for the single pulse train in the satellite unstable regime.
The pulse intensity for $E$ (blue) and $Y$ (orange) fields are shown.
With realistic values of the linewidth enhancement factors the dynamics
is more involved due to the chirp that induces pulse broadening. Here,
instead of dying out completely the parent pulse merges with its growing
satellite. Parameters are $(J_{1},\alpha_{1},\alpha_{2})=(0.65,2.1,0.5)$. }
\label{fig:with_LEFs} 
\end{figure}

With realistic linewidth enhancement factors for semiconductor media
the satellite and self-phase modulation instabilities combine to form
a dynamics of the kind presented in \citep{SCM_PRL_19}. By performing
parameter scans as a function of various parameters we were able to
deduce some rules of thumb for this satellite instability. Generally,
the pulse-width is proportional to the photon lifetime and can be
written as $\tau_{p}=\kappa^{-1}f\left(\cdots\right)$ with $f\left(\cdots\right)$
a function that depends on all the other parameters of the PML setup.
Optimizing PML consists in finding the parameter combination for which
the function $f$ is the smallest. Since the temporal separation between 
the main pulse and its first sattelite is fixed by the cavity photon lifetime, 
it is \emph{in these optimal cases} where, as depicted in Fig.~\ref{fig:clean_sat}, 
that the satellites become better resolved from the main pulse and where they are prone to become
unstable. Hence, the satellite instability can be obtained by tuning
any parameter leading to the optimal PML and shortest pulse-width,
making this instability an essential limitation in the optimization
procedure to find the narrowest pulses. For instance, increasing
the saturation $s$ or even the linewidth enhancement factor of the
absorber $\alpha_{2}$, if it compensates for the chirp induced in
the gain section, can adversely narrow the pulse which destabilizes the PML regime.
In "bad conditions" the pulse and its sattelites are smeared out into a globally broader pulse that is stable.
Note that in the VCSEL-SESAM setup studied in \citep{SCM_PRL_19} the detuning between
the two micro-cavities is an additional factor leading to further
complexity since second order dispersion also plays a role. 
Finally, we show in Fig.~\ref{fig:with_LEFs} how the
satellite instability evolves for more realistic situations settubg $\alpha_{1}=2.1$
and $\alpha_{2}=0.5$. We note that, in addition, this regime can
be bistable with the stable pulsating solution, 
as shown in the orange region in Fig.~\ref{fig:bif_diag}(a).

\section{The Dispersive Master Equation}

\label{sec:pde}

Bifurcation analysis does not always allows for an intuitive interpretation
of the dynamics. In addition, even if the time delayed equations Eqs.~(\ref{eq:DAE1}-\ref{eq:DAE4})
contain a lot of the physics of PML, their form remains complicated.
For instance, their dispersive features do not appear so obvious.
We derive in this section a PDE for the field amplitude $E$ that
approximates the dynamics of the full DADE model~(\ref{eq:DAE1}-\ref{eq:DAE4}).
In the PDE representation, the field depends on a slow and a fast
time, i.e. $E\equiv E\left(\xi,z\right)$. Here, the slow time $\left(\xi\right)$
represents the evolution of the field profile from one round-trip
to the next while the ``spatial variable'' $\left(z\right)$ describes
the fast evolution of the pulse within the round-trip. While this
approach has been used in several time-delayed systems, see e.g.,
\citep{YG-JPA-17,EJW-CHA-17} for a review, such PDE models are usually
termed Haus master equations in the framework of mode-locking, see
\citep{haus00rev} for a review.

In the case of high frequency PML dynamics, i.e. the regimes in which
the cavity round-trip is much shorter than the gain recovery time,
a PDE approximating the dynamics of a ring cavity model based upon
delay differential equations \citep{VT-PRA-05} was proposed in \citep{KNE-PD-06}.
The multiple time-scales analysis method was used and the scaling
of parameters consisted in assuming low losses, low gain, and weak
spectral filtering. In the model of \citep{VT-PRA-05}, these three
physical effects are controlled by three independent parameters. In
the multiple time-scales approach one finds, at the lowest order,
a periodic solution, e.g., a pulse evolving over the fast time scale
$\left(z\right)$ that circulates in the cavity without deformation.
At third order in the expansion scheme, a solvability condition allows
finding that the dynamics on the slow time scale $\left(\xi\right)$
is governed by the weak effects of gain, loss and spectral filtering.
In a PDE representation, the gain filtering in the model of \citep{VT-PRA-05}
takes the form of a diffusion over the fast time, i.e. a term $d_{2}\partial_{z}^{2}E$
with $d_{2}>0$. While the smallness of gain, losses and filtering
can be, in some situations, debatable, the advantage of the approach
presented in \citep{KNE-PD-06} is the uniform accuracy of the PDE
representation that was not, e.g., limited to the vicinity of the
lasing threshold. For instance, no \emph{a priori} conditions over
the magnitude of the field were necessary.

It is interesting ---and surprising--- to notice that the aforementioned
approach fails if one tries to export it to the case of the DADE model~(\ref{eq:DAE1}-\ref{eq:DAE4});
the resulting PDE obtained similarly as a third order solvability
condition does possess gain and losses, yet it is devoid of spectral
filtering $\left(d_{2}=0\right)$ which leads to singular dynamics
and unphysical pulse collapse. The physical reason that underlies
this mathematical phenomenon is that the filtering of the micro-cavity
is not in our modeling approach independent from gain and losses.
It is the actual level of the population inversion in the micro-cavity
that defines the breadth and hence the curvature of the resonance.
This effect is particularly pronounced in the Gires-Tournois regime
(i.e., $h=2$), where the empty cavity reflectivity is unity, which
corresponds to no curvature at all and $d_{2}=0$. That is, positive
(resp. negative) curvature induces diffusion (resp. anti-diffusion).
Generally, the average gain experienced by the pulse must be positive
to compensate for the cavity losses incurred by the mirror reflectivity
$\eta$.

A proper analysis of the model given by Eqs.~(\ref{eq:DAE1}-\ref{eq:DAE4})
shall result in a PDE whose diffusion term $d_{2}$ depends on other
parameters such as the cavity losses. This discussion materializes
by taking as the expansion point the cavity at the lasing threshold
instead of the empty cavity, as done in \citep{KNE-PD-06}. At the
lasing threshold, the unsaturated gain and losses exactly compensate.
This modification will allow to obtain the filtering induced by the
cavity at threshold, instead of that of the empty cavity leading to
$d_{2}=0$. The drawback of our approach is that we have to assume
the pulse to be not too intense and treat the nonlinear effects pertubatively,
limiting our analysis to the vicinity of the lasing threshold.

We start by normalizing time by the cavity round-trip $\tau$ time
as $\sigma=t/\tau$ and define a smallness parameter $\varepsilon=1/\tau$.
As we operate in the long cavity limit, the carriers are not independent
functions of time that can lead to resonant terms and solvability
conditions. Instead, the carrier evolutions depend uniquely on the
initial conditions at the beginning of the round-trip, which in the
long cavity limit is the equilibrium value, and on the amplitude of
the field, i.e., $N_{j}=N_{j}\left(J_{j},E\right)$. Hence, we can
concentrate solely on the field dynamics that reads 
\begin{eqnarray}
\varepsilon\frac{dE}{d\sigma} & = & \left[\left(1-i\alpha_{1}\right)N_{1}+\left(1-i\alpha_{2}\right)N_{2}-1\right]E+hY,\\
Y\left(\sigma\right) & = & \eta\left[E\left(\sigma-1\right)-Y\left(\sigma-1\right)\right]\,.
\end{eqnarray}
We assume a small deviation of the gain and absorber with respect
to their equilibrium values that we scale as 
\begin{eqnarray}
N_{j} & = & J_{j}+\varepsilon^{3}n_{j}
\end{eqnarray}
with $j\in\left[1,2\right]$. We also assume $\eta$ to be real as
the feedback case is irrelevant in the long cavity regime. Defining
the Fourier transform of the field profiles at the $n$-th round-trip
as $\left(E_{n},Y_{n}\right)$ and using that $\dfrac{d}{d\sigma}\rightarrow-i\omega$
we obtain 
\[
\left(1-G_{t}-i\varepsilon\omega\right)E_{n}=hY_{n}+\varepsilon^{3}\sum_{j=1}^{2}\left(1-i\alpha_{j}\right)\left(n_{j}E\right)_{n},
\]
where we used the shorthand for the total complex gain~$G_{t}=\left(1-i\alpha_{1}\right)J_{1}+\left(1-i\alpha_{2}\right)J_{2}.$
Noticing that the DAE for $Y\left(\sigma\right)$ in Fourier space
reads 
\begin{equation}
Y_{n}+\eta Y_{n-1}=\eta E_{n-1}
\end{equation}
and, by making a linear combination of $E_{n}$ and $\eta E_{n-1}$,
we get after simplification \emph{a functional mapping} 
\begin{eqnarray}
E_{n} & = & \eta\frac{h-1+G_{t}+i\varepsilon\omega}{1-G_{t}-i\varepsilon\omega}E_{n-1}\label{eq:Map_ini}\\
 & + & \varepsilon^{3}\frac{1}{1-G_{t}-i\varepsilon\omega}\sum_{j=1}^{2}\left(1-i\alpha_{j}\right)\left[\left(n_{j}E\right)_{n}+\eta\left(n_{j}E\right)_{n-1}\right].\nonumber 
\end{eqnarray}

Now we impose the value of $G_{t}$ to be a convenient expansion point
and we set the threshold condition over the linear multiplier $\mu$
\begin{eqnarray}
\mu=\eta\frac{h-1+G_{t}+i\varepsilon\omega}{1-G_{t}-i\varepsilon\omega} & = & 1,
\end{eqnarray}
which allows finding the lasing frequency shift at threshold as $\varepsilon\omega_{t}=\alpha_{1}J_{1}+\alpha_{2}J_{2}$.
This leaves us with a real equation for the amplification factor 
\begin{eqnarray}
\eta\frac{h-1+N_{t}}{1-N_{t}} & = & 1\,,
\end{eqnarray}
where we defined $N_{t}=J_{1}+J_{2}$. The last relation implies that
the threshold is defined by 
\begin{eqnarray}
N_{t}=J_{1}+J_{2} & = & 1-\frac{h\eta}{1+\eta}\,.\label{eq:Nth}
\end{eqnarray}
We can now express the field multiplier $\mu$ from one round-trip
towards the next as 
\begin{eqnarray}
\mu\left(\omega,N_{t}\right) & = & \eta\frac{h-1+N_{t}+i\varepsilon\left(\omega-\omega_{t}\right)}{1-N_{t}-i\varepsilon\left(\omega-\omega_{t}\right)}
\end{eqnarray}
and the functional mapping given by Eq.~\ref{eq:Map_ini} reads
\begin{widetext}
\begin{eqnarray}
E_{n} & = & \eta\frac{h-1+N_{t}+i\varepsilon\left(\omega-\omega_{t}\right)}{1-N_{t}-i\varepsilon\left(\omega-\omega_{t}\right)}E_{n-1}+\varepsilon^{3}\frac{1}{1-N_{t}-i\varepsilon\left(\omega-\omega_{t}\right)}\sum_{j=1}^{2}\left(1-i\alpha_{j}\right)\left[\left(n_{j}E\right)_{n}+\eta\left(n_{j}E\right)_{n-1}\right]\,.\label{eq:Map_inter}
\end{eqnarray}
\end{widetext}

One only needs Eq.~\ref{eq:Map_inter} to be accurate up to third
order in $\varepsilon$ in order to obtain the proper expression of
the diffusion, TOD and nonlinear terms. As such, we can simplify the
last term of Eq.~\ref{eq:Map_inter} by replacing the value of $E_{n}$
at the lowest order, i.e., we can set 
\begin{equation}
E_{n}=\eta\frac{h-1+N_{t}}{1-N_{t}}E_{n-1}+\mathcal{O}\left(\varepsilon\right)=E_{n-1}+\mathcal{O}\left(\varepsilon\right),
\end{equation}
where we used the threshold definition given by Eq.~\ref{eq:Nth}.
Using that $E_{n}=E_{n-1}$ in the nonlinear term of~\ref{eq:Map_inter},
replacing the expression of the threshold and noticing that all the
frequencies are relative to that of the lasing threshold, so that
one can set $\tilde{\omega}=\omega-\omega_{t}$, yields the expression
\begin{eqnarray}
E_{n} & = & \eta\frac{h-1+N_{t}+i\varepsilon\tilde{\omega}}{1-N_{t}-i\varepsilon\tilde{\omega}}E_{n-1}\label{eq:MapSimple}\\
 & + & \varepsilon^{3}\frac{\left(1+\eta\right)^{2}}{h\eta}\sum_{j=1}^{2}\left(1-i\alpha_{j}\right)\left(n_{j}E\right)_{n-1}+\mathcal{O}\left(\varepsilon^{4}\right)\,.\nonumber 
\end{eqnarray}

Now let us assume there exists a PDE for the field $E\left(\xi,z\right)$
with $\xi$ and $z$ the slow and fast times, respectively. In Fourier
space for the variable $z$, one obtains 
\begin{eqnarray}
\partial_{\xi}E & = & \mathcal{L}\left(\tilde{\omega}\right)E+\varepsilon^{3}\mathcal{N}\left(\xi,\tilde{\omega}\right)\,.\label{eq:PDE_ansatz}
\end{eqnarray}
The form of Eq.~\ref{eq:PDE_ansatz} consists naturally of a linear
operator $\mathcal{L}\left(\tilde{\omega}\right)$ that should correspond
to the linear multiplier of the mapping in Eq.~\ref{eq:MapSimple},
while $\mathcal{N}\left(\sigma,\tilde{\omega}\right)$ accounts for
nonlinear gain and absorber effects. We assume that $\mathcal{L}\left(\tilde{\omega}\right)$
is small, i.e, $\mathcal{L}\left(\omega\right)\sim0+\mathcal{O}\left(\varepsilon\right)$.
Such a scaling is consistent with the definition of the lasing threshold
and will be checked a posteriori. Integrating \emph{exactly} Eq.~\ref{eq:PDE_ansatz}
over a round-trip yields 
\begin{eqnarray}
E_{n} & = & e^{\mathcal{L}}E_{n-1}+\varepsilon^{3}\int_{n-1}^{n}e^{\left(n-\xi\right)\mathcal{L}}\mathcal{N}\left(\xi,\tilde{\omega}\right)d\xi\,.\label{eq:PDE2MAP}
\end{eqnarray}
Because the integral term in Eq.~\ref{eq:PDE2MAP} is already at
third order in $\varepsilon$, we can approximate $e^{\left(n-\xi\right)\mathcal{L}}=1+\mathcal{O}\left(\varepsilon\right)$
and evaluate the nonlinear operator using the Euler explicit method.
Indeed, since $\mathcal{N}$ depends on the field, the error in the
integration will be proportional to the slow evolution of the field
from one round-trip towards the next, i.e., $\partial_{\xi}\mathcal{N}\sim\partial_{\xi}E\sim\mathcal{O}\left(\varepsilon\right)$.
That is, we find 
\begin{eqnarray}
E_{n} & = & e^{\mathcal{L}}E_{n-1}+\varepsilon^{3}\mathcal{N}\left(n-1,\tilde{\omega}\right)+\mathcal{O}\left(\varepsilon^{4}\right).\label{eq:PDE_sol}
\end{eqnarray}
Comparing Eq.~\ref{eq:MapSimple} and Eq.~\ref{eq:PDE_sol} we deduce
that 
\begin{eqnarray}
\mathcal{L} & = & \ln\left(\eta\frac{h-1+N_{t}+i\varepsilon\tilde{\omega}}{1-N_{t}-i\varepsilon\tilde{\omega}}\right)\,.
\end{eqnarray}
Using Eq.~\ref{eq:Nth} the last expression can be simplifies as
\begin{eqnarray}
\mathcal{L} & = & \ln\left(\frac{1+i\varepsilon\eta\frac{1+\eta}{h\eta}\tilde{\omega}}{1-i\varepsilon\frac{1+\eta}{h\eta}\tilde{\omega}}\right)\,.\label{eq:goodL}
\end{eqnarray}
We can verify easily that $\mathcal{L}=\mathcal{O}\left(\tilde{\varepsilon}\right)$
which allows to check a posteriori our approximation regarding the
order of the operator $\mathcal{L}$. One can also expand Eq.~\ref{eq:goodL}
in $\tilde{\omega}$ up to third order will yields the drift, diffusion
and TOD coefficients $d_{1},d_{2}$ and $d_{3}$ as 
\begin{equation}
\mathcal{L}=d_{1}\left(-i\varepsilon\tilde{\omega}\right)+d_{2}\left(-\varepsilon^{2}\tilde{\omega}^{2}\right)+d_{3}\left(i\varepsilon^{3}\tilde{\omega}^{3}\right)+\mathcal{O}\left(\varepsilon^{4}\right)
\end{equation}
with 
\begin{eqnarray}
d_{1} & = & -\frac{\left(\eta+1\right)^{2}}{h\eta}\,,\\
d_{2} & = & \frac{1-\eta^{2}}{2}\left(\frac{\eta+1}{h\eta}\right)^{2}\,,\\
d_{3} & = & -\frac{\eta^{3}+1}{3}\left(\frac{\eta+1}{h\eta}\right)^{3}\,.
\end{eqnarray}
The values of the coefficient $d_{j}$ is particularly instructive
and, in particular, how they deviate from the expression one can find
easily in the case of an empty Gires-Tournois micro-cavity coupled
to an external mirror: $(d_{1},d_{2},d_{3})=(-2,0,-2/3)$. Here, we
notice that although $d_{1}$ and $d_{3}$ are somehow modified by
the value of the cavity losses, the value of $d_{2}$ is not vanishing.
In the good cavity limit $\eta\rightarrow1$, we have $d_{2}\sim1-\eta\rightarrow0$,
which explains why TOD is important as it becomes the leading order
term.

Further, we identify the nonlinear operator $\mathcal{N}$ as 
\begin{eqnarray}
\mathcal{N}\left(n-1,\tilde{\omega}\right) & = & \frac{\left(1+\eta\right)^{2}}{h\eta}\sum_{j=1}^{2}\left(1-i\alpha_{j}\right)\left(n_{j}E\right)_{n-1}
\end{eqnarray}
Finally, reverting Eq.~\ref{eq:PDE_ansatz} to direct space using
that $-i\varepsilon\tilde{\omega}\rightarrow\partial_{z}$, the sought
PDE for the field $E$ reads 
\begin{eqnarray}
\partial_{\xi}E & = & \left(d_{1}\partial_{z}+d_{2}\partial_{z}^{2}+d_{3}\partial_{z}^{3}\right)E\label{eq:PDE_qfinal}\\
 & + & \frac{\left(1+\eta\right)^{2}}{h\eta}\sum_{j=1}^{2}\left(1-i\alpha_{j}\right)\left(N_{j}-J_{j}\right)E\,.\nonumber 
\end{eqnarray}
By using the definition of the lasing threshold and of the carrier
frequency, we find that the dispersive master equation for the field
$E$ in the long cavity limit reads
\begin{widetext}
\begin{eqnarray}
\partial_{\xi}E & = & \left(d_{1}\partial_{z}+d_{2}\partial_{z}^{2}+d_{3}\partial_{z}^{3}\right)E+\frac{\left(1+\eta\right)^{2}}{h\eta}\left\{ \left(1-i\alpha_{1}\right)N_{1}+\left(1-i\alpha_{2}\right)N_{2}-1+\frac{h\eta}{1+\eta}-i\omega_{t}\right\} E\,,\label{eq:PDE_final_E}
\end{eqnarray}
\end{widetext}
whereas the equations for the carriers take the form 
\begin{eqnarray}
\frac{\partial N_{1}}{\partial z} & = & \gamma_{1}(J_{1}-N_{1})-|E|^{2}N_{1}\,,\label{eq:PDE_final_N1}\\
\frac{\partial N_{2}}{\partial z} & = & \gamma_{2}(J_{2}-N_{2})-s|E|^{2}N_{2}\,.\label{eq:PDE_final_N2}
\end{eqnarray}
Note that the rotation term $i\omega_{t}$ in Eq.\eqref{eq:PDE_final_E}
is immaterial and can be removed by setting $\tilde{E}\left(\xi,z\right)=E\left(\xi,z\right)\exp\left(i\omega_{t}\xi\right)$.

    \begin{figure}[ht]
    \includegraphics[viewport=60bp 10bp 1100bp 740bp,clip,width=0.49\textwidth]{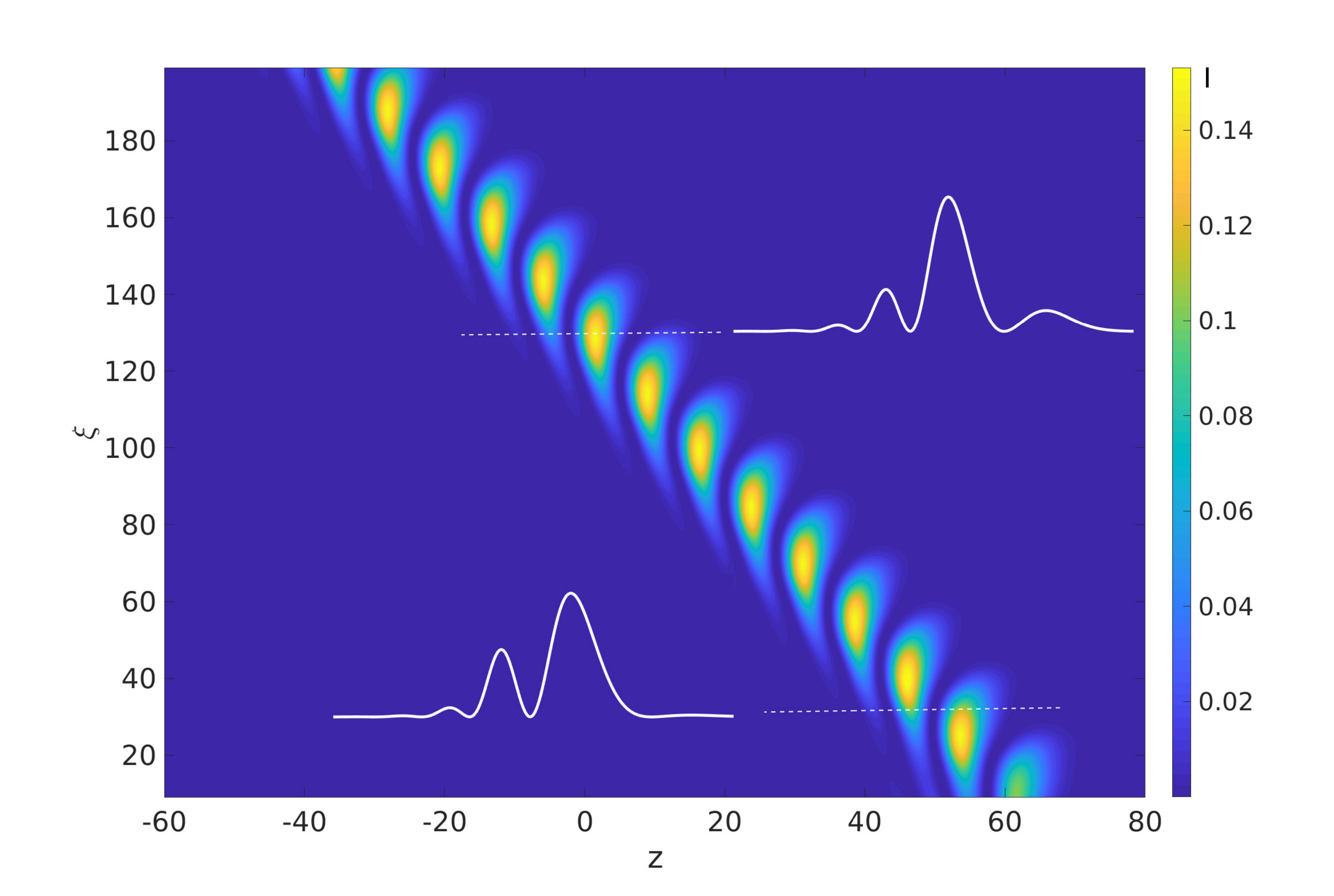}
        \caption{(color online)
            Space-time diagram of the satellite instability found in the DNS of the master PDE (\ref{eq:PDE_final_E}-\ref{eq:PDE_final_N2}). The pulse intensity $I=|E|^2$ is shown. Parameters are $(J_1,\,J_2,\,\alpha_1,\,\alpha_2,\,\eta,\,s)=(0.119,\,-0.1,\,0,\,0,\,0.9,\,15)$.}
        \label{fig:PDE_sat}
    \end{figure}

First we studied the PDE model~\eqref{eq:PDE_final_E}-\eqref{eq:PDE_final_N2}
numerically to demonstrate the existence of the satellite instability
in the parameter space and the resulting space-time diagram obtained
for zero linewidth enhancement factors $\alpha_{1}=\alpha_{2}=0$
is shown in Fig.~\ref{fig:PDE_sat}. Again, we observe the clean
cut satellite instability in this parameter range, but also that our PDE 
approximates very well the dynamics of the underlying time delayed model.

\begin{figure}[ht]
        \includegraphics[width=.24\textwidth]{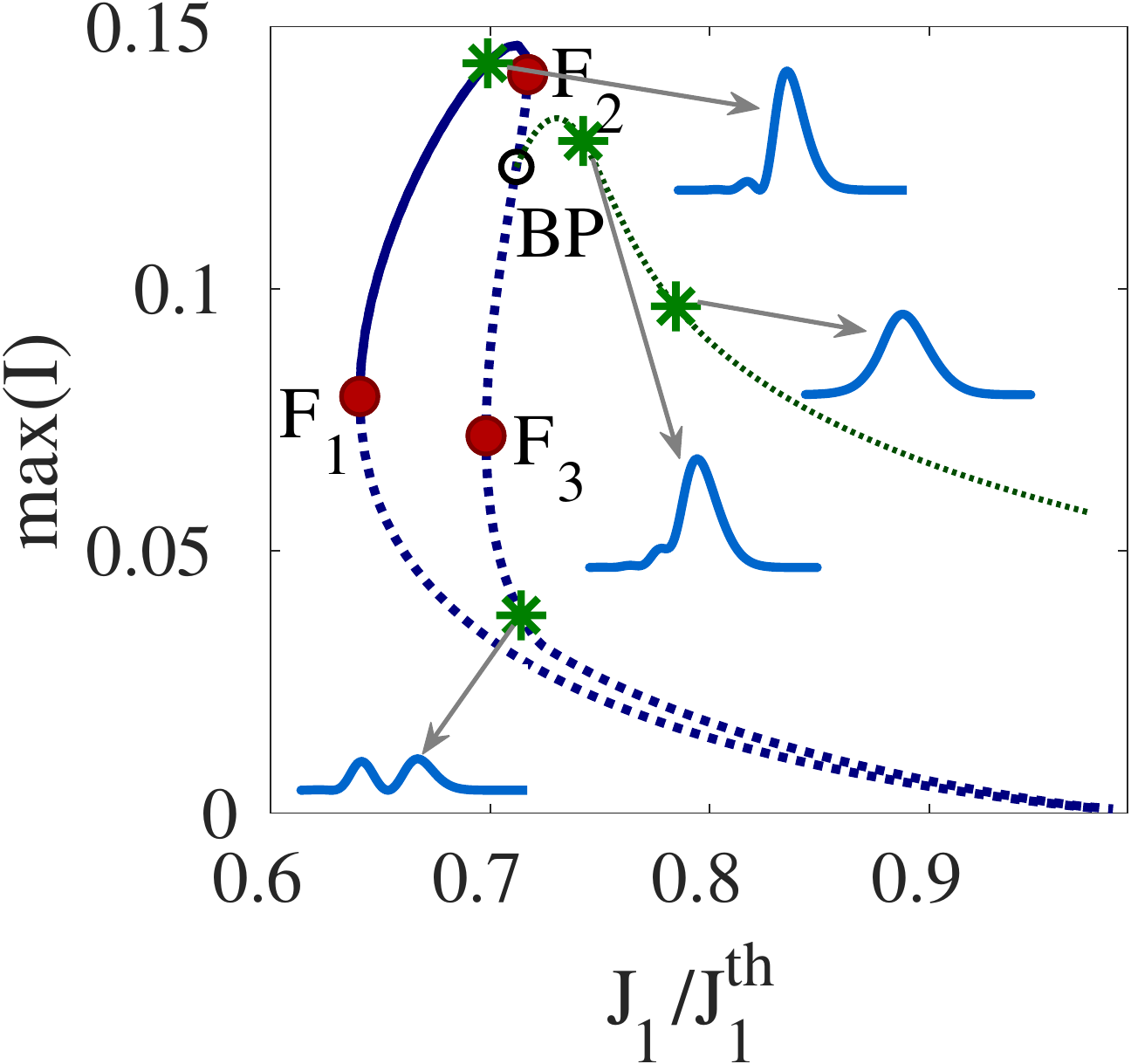}
        \includegraphics[width=.23\textwidth]{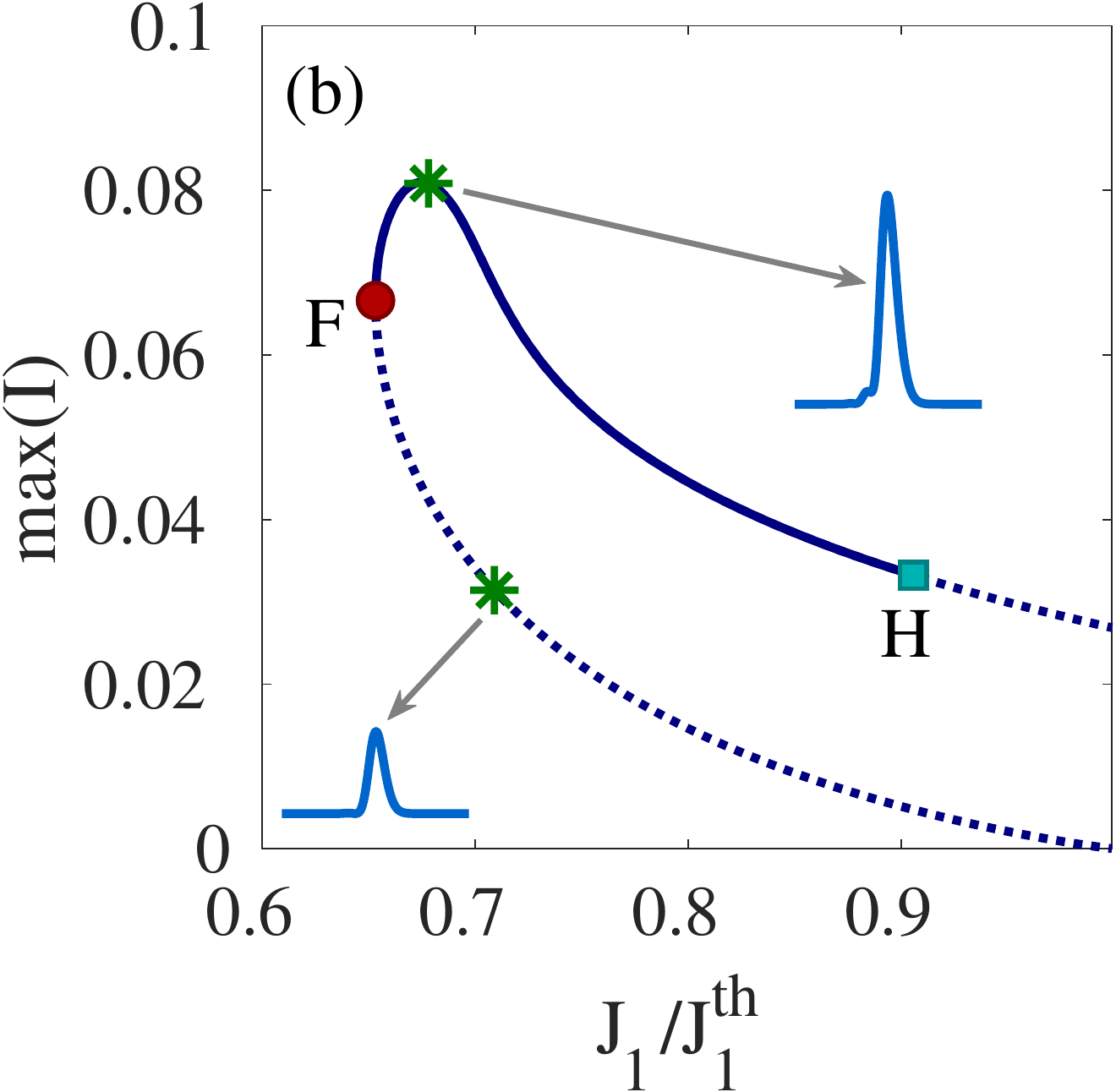}
        \caption{(color online)
            Branches of the TLSs in the master PDE~\eqref{eq:PDE_final_E}-\eqref{eq:PDE_final_N2}. The maximum of the field intensity $I=|E|^2$ as a function of the normalized pump current is shown. (a) For $(\alpha_1,\,\alpha_2)=(0,0)$ the pulse profiles show defined satellites. The branch has three folds $\mathrm{\bold F}_i$ and a branching point $\mathrm{\bold{BP}}$. The second fold $\mathrm{\bold F}_2$ coincides with an infinite period limit cycle in a global SNIPER bifurcation. An unstable branch emerges from $\mathrm{\bold{BP}}$ in a pitchfork bifurcation. (b) For $(\alpha_1,\,\alpha_2)=(1.5,0.5)$, $\mathrm{\bold F}_2$ and $\mathrm{\bold{BP}}$ have merged into an AH bifurcation point $\mathrm{\bold H}$.  The profiles only show intensity bumps on the leading edge of the pulse and there is no longer a SNIPER.  Other parameters are $(J_2,\,\eta,\,s)=(-0.1,\,0.9,\,15)$.
        }
        \label{fig:PDE_branches}
\end{figure}

To compare the PDE~\eqref{eq:PDE_final_E}-\eqref{eq:PDE_final_N2}
with the DAE~(\ref{eq:DAE1}-\ref{eq:DAE4}) in detail we performed
bifurcation analysis of the PDE by using pseudo-arclength continuation
methods within the pde2path framework~\citep{uecker2014}. To this
aim, first we seek for the steady localized pulse solutions of Eqs.~\eqref{eq:PDE_final_E}-\eqref{eq:PDE_final_N2}
that can be found by setting $E(z,\xi)=A(z-\upsilon\xi)e^{-i\varpi\xi}$
leading to the following equation for the stationary field $A$ 
\begin{align}
0 & =\upsilon\,\frac{\partial A}{\partial z}+d_{2}\frac{\partial^{2}A}{\partial z^{2}}+d_{3}\frac{\partial^{3}A}{\partial z^{3}}+i\varpi\,A\\
 & \quad+\frac{(1+\eta)^{2}}{h\eta}\left((1-i\alpha)N_{1}+(1-i\beta)N_{2}-1+\frac{h\eta}{1+\eta}\right)A\,.\nonumber 
\end{align}
Note that both the spectral parameter $\varpi$ and the drift velocity
$\upsilon$ become free parameters that can be found by imposing additional
auxiliary integral conditions. In addition we set the following boundary
conditions for the domain $z\in[0,L]$ 
\begin{eqnarray*}
\frac{\partial A}{dz}\Big|_{z=0,L} & = & 0\,,\\
N_{1}\big|_{z=0} & = & J_{1}\,,\quad-\frac{\partial N_{1}}{\partial z}\Big|_{z=L}+\gamma_{1}\left(J_{1}-N_{1}\big|_{z=L}\right)=0\,,\\
N_{2}\big|_{z=0} & = & J_{2}\,,\quad-\frac{\partial N_{1}}{\partial z}\Big|_{z=L}+\gamma_{2}\left(J_{2}-N_{2}\big|_{z=L}\right)=0\,.
\end{eqnarray*}
Now we can follow the TLS of the PDE ~\eqref{eq:PDE_final_E}-\eqref{eq:PDE_final_N2}
in parameter space and in Fig.~\ref{fig:PDE_branches} we present
two branches of TLSs for different values of the linewidth enhancement
factors. Panel (a) shows the intensity of the TLS as a function of
the normalized pump rate for the case of $\alpha_{1}=\alpha_{2}=0$.
Like in the DAE case (cf. Fig.\ref{fig:bif_diag}), one can see that
the branch folds three times (points $\mathrm{\bold F}_{i}$) when
continuing in the pump rate $J_{1}$ and the second fold $\mathrm{\bold F}_{2}$
is responsible for the SNIPER bifurcation after the first leading
satellite becoming sufficiently large to saturate the absorber (cf.
Fig. \ref{fig:PDE_sat}). Note that an additional unstable branch
connects to the main one in a branching point $\mathrm{\bold{BP}}$.
Panel (b) shows the same gain interval for non-vanishing linewidth
enhancement factors $\alpha_{1}=1.5$ and $\alpha_{2}=0.5$. Here,
the branch continues without the additional folds $\mathrm{\bold F}_{2}$
and $\mathrm{\bold F}_{3}$ and only becomes Andronov-Hopf unstable
at large gain value. Indeed, the second fold and the branching point
have merged while the part of the branch with the third fold has detached,
giving a qualitatively different scenario. The chirp induced by the
linewidth enhancement factors smears out the TOD effect responsible
for the satellites.

\begin{figure}[ht]
        \includegraphics[height=5cm]{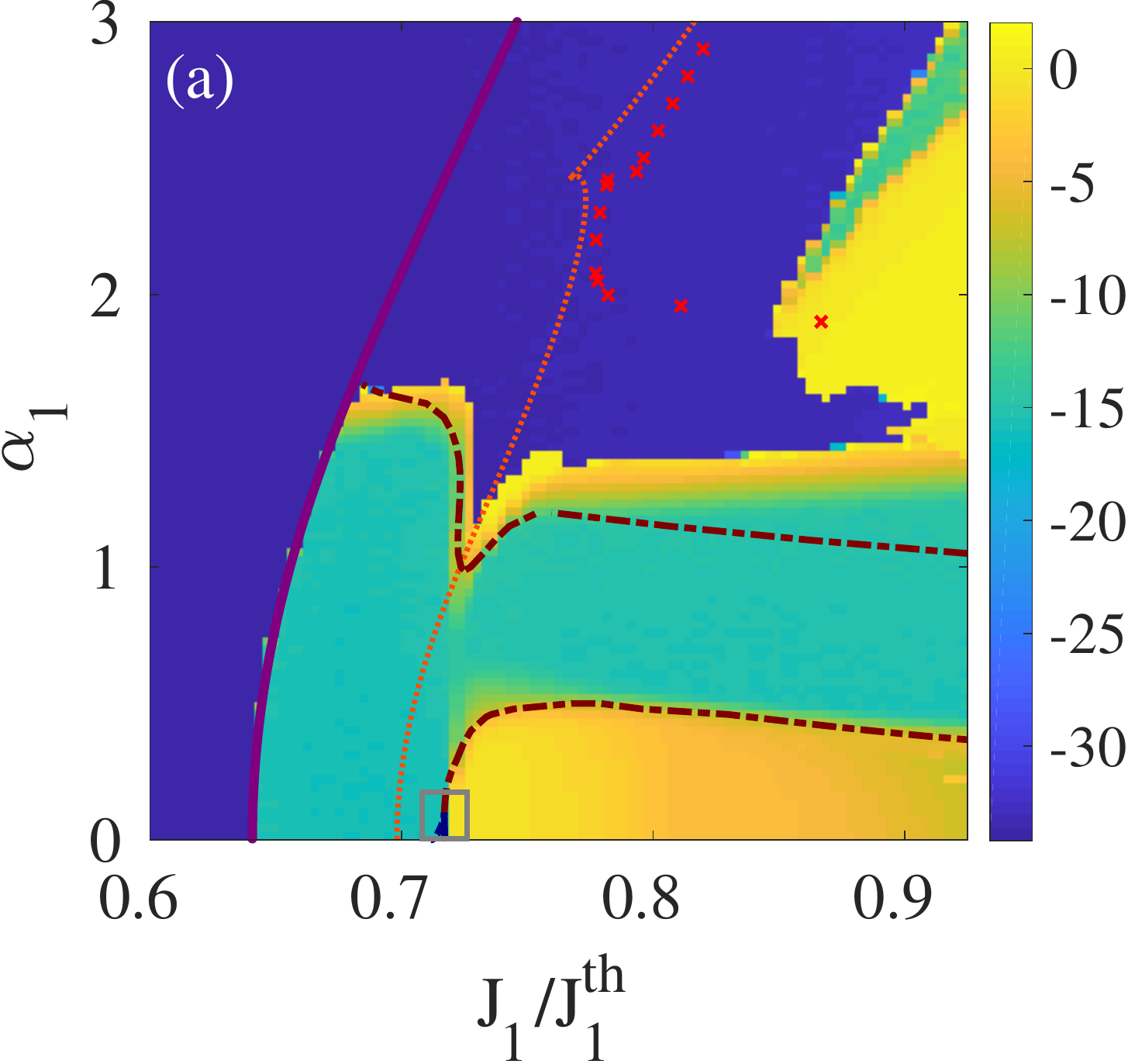}
        \hspace{0.1cm}
        \includegraphics[height=5cm]{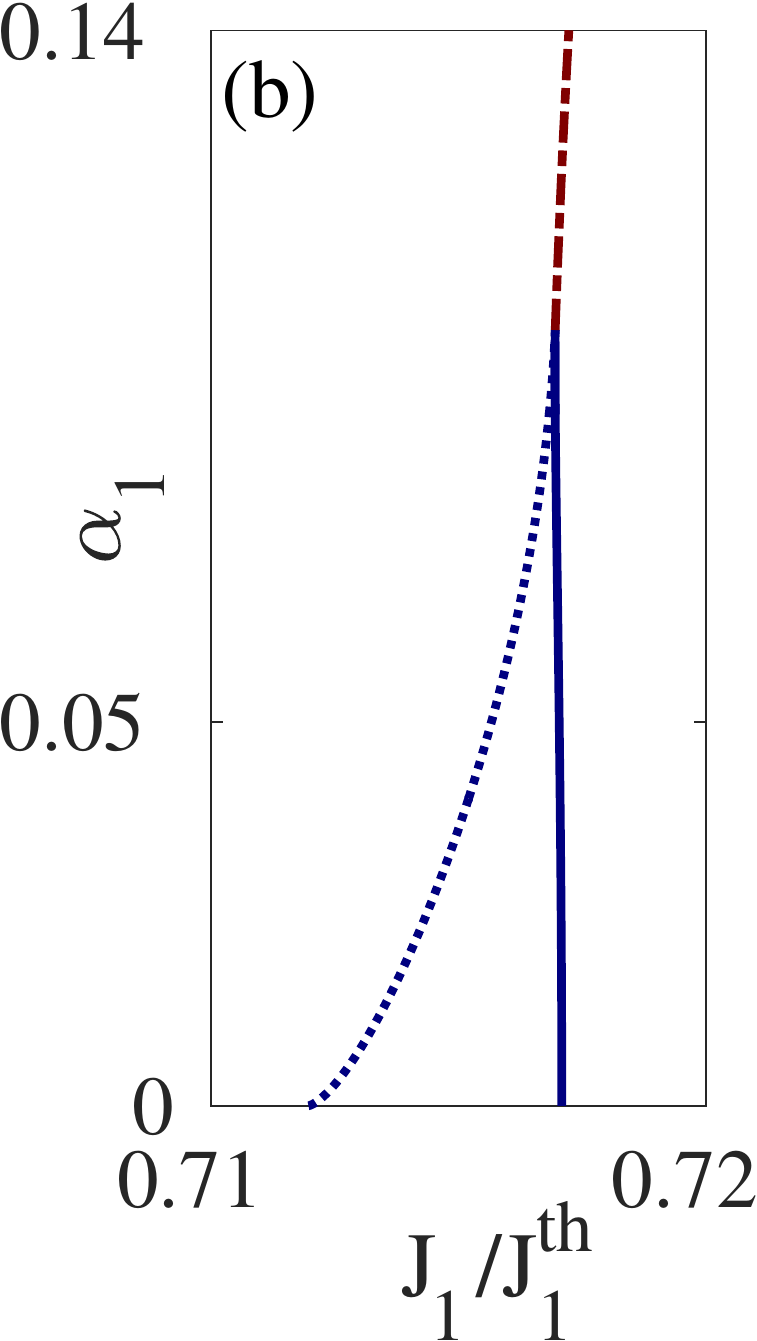}
        \caption{(color online)
            (a) Bifurcation diagram in the $(J_1, \alpha_1)$ plane of the DAE model~(\ref{eq:DAE1}-\ref{eq:DAE4}) in the long delay limit superposed with the bifurcation diagram of the equivalent PDE~\eqref{eq:PDE_final_E}-\eqref{eq:PDE_final_N2}. The color coding shows the standard deviation of the pulse energy obtained by DNSs of the DAE. The evolution of the fold $\mathrm{\bold F}_1$ is marked by a solid red line, the satellite instability around $\alpha_1=0$ with the corresponding fold $\mathrm{\bold F}_3$  (branching point) is in solid (dotted) blue and the AH part in dash dotted red. The other AH bifurcation corresponding to self-phase modulation slopes down for the PDE case in contrast to the DAE. The high $\alpha_1$ region in the PDE is significantly shifted with respect to the DAE.  Its fold branch is depicted in dotted orange and AH bifurcations are indicated by red crosses.  Parameters are $(J_2,\,\alpha_2,\,\eta,\,s)=(-0.1,\,0,\,0.9,\,15)$.  (b) Zoom-in on the SNIPER region where a fold and a branching point merge into an AH bifurcation.
        }
        \label{fig:PDE_compare}
    \end{figure}

Finally, in Fig.~\ref{fig:PDE_compare} we superpose the results
on top of data obtained through DNSs of the DAE model~(\ref{eq:DAE1}-\ref{eq:DAE4})
in the long delay limit using the functional mapping approach \citep{SJG-OL-18}.
Pulses are fully localized TLSs in this regime. In panel (a) the standard
deviation of the pulse energy is shown as a function of the gain bias
$J_{1}$ normalized to threshold $J_{1}^{th}$ and the gain linewidth
enhancement factor $\alpha_{1}$ along with the bifurcation curves
from continuation. For small $\alpha_{1}$ one can see the satellite
unstable region which is similar to the previous parameter set, i.e.,
close to $\alpha_{1}=0$ there is an additional fold presented in
solid blue and the satellite instability sets in as a global SNIPER
bifurcation after it. Panel (b) shows a zoom-in on this area where
the fold merges with a branching point (dotted blue) thereby forming
an AH bifurcation depicted in dash dotted red. Both models quantitatively
agree in this area. The principal fold of the subcritical TLS branch
in solid red is also reproduced correctly. For higher $\alpha_{1}$
the stable pulse region is limited by another AH bifurcation corresponding
to self-phase modulation, shown as well in dash dotted red. In contrast
to the DAE, for the PDE model the bifurcation curve slopes down in
$\alpha_{1}$ for increasing gain. Both the nature of the instability
and the discrepancy found in the equivalent PDE are somewhat similar
to the bifurcation structure in the Vladimirov-Turaev model for passive
mode-locking in a unidirectional ring laser \citep{VT-PRA-05,SJG-PRA-18}.

\begin{figure}
        \includegraphics[viewport=10bp 0bp 470bp 420bp,clip,width=0.49\textwidth]{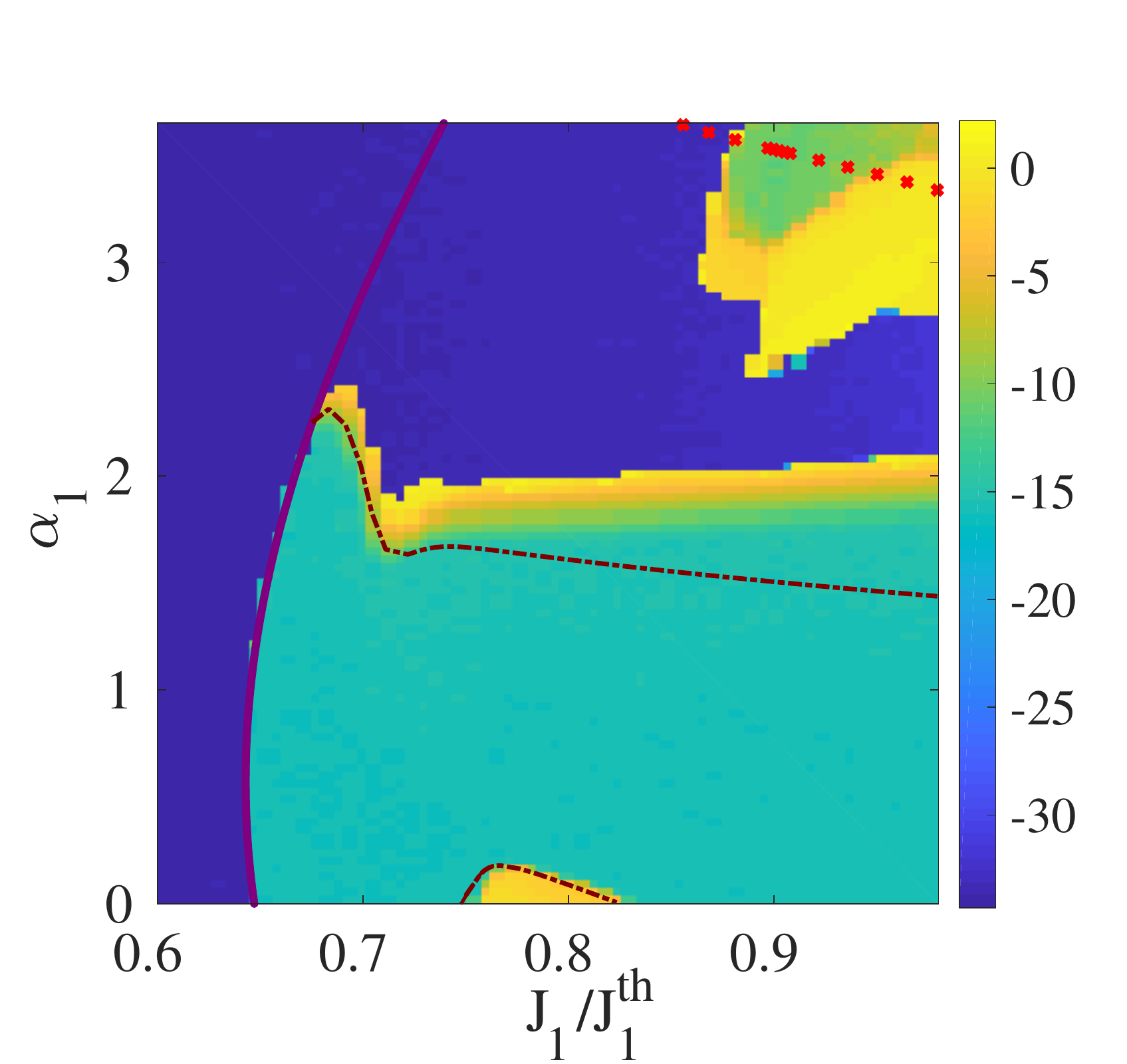}
        \caption{(color online)
            For larger $\alpha_2$ the different regions move and deform. The stable pulse region moves up in $\alpha_1$ by a similar amount while the satellite region moves down. The high $\alpha_1$ region completely detaches in the area below threshold.  Results from bifurcation analysis of the PDE strongly differs from the DAE in this region. Parameters are $(J_2,\,\alpha_2,\,\eta,\,s)=(-0.1,\,0.5,\,0.9,\,15)$.
        }
        \label{fig:PDE_compare_05}
    \end{figure}

For the parameters of Fig~\ref{fig:PDE_compare}(a) the DAE system
exhibits a region close to threshold at high values of $\alpha_{1}$
that is partly stable on the high $\alpha_{1}$ edge. The corresponding
fold and AH curves shown in dotted orange or indicated by red crosses,
respectively, are found in the PDE with a qualitatively similar shape
but the position of this area is shifted significantly towards lower
gain as compared to the DAE. This region corresponds to the bistable
region for the previous parameter set. Indeed we found that generally
it moves and changes shape significantly as a function of the other
parameters and so do the principal pulse and satellite instability
regions. For example increasing $\alpha_{2}$ makes the stable pulses
follow by moving up in $\alpha_{1}$ by roughly the same amount, while
the satellite instability moves down. The second region moves up in
$\alpha_{1}$ much quicker and completely detaches, at least when
constricting one's view at the area below threshold (see Fig.~\ref{fig:PDE_compare_05}).

\section{Conclusion}

In conclusion, we discussed the dynamics of MIXSELs using a first-principle dynamical model based upon delay algebraic equations. We have found that the third order dispersion induced by the micro-cavity induces satellites on the leading edge of the pulses and shown that the latter can become unstable.

Using a combination of direct time simulations and a path-continuation methods we reconstructed the branch of a single pulse solution. In the limit of vanishing line enhancement factors we show that the onset of the satellite instability is associated with a global bifurcation of the saddle-node infinite period type. The influences of noise as well as features of excitability are discussed.

In the case of non-vanishing linewidth enhancement factors, we showed that the satellite and self-phase modulation instabilities can combine leading to a intricate oscillating dynamics which shed further light on the results obtained in a VCSEL-SESAM setup \citep{SCM_PRL_19}.

Finally, we derived an approximate dispersive master PDE model and performed
a full bifurcation analysis. We demonstrated that this PDE reproduces
the satellite as well as the low $\alpha_{1}$ SNIPER and AH bifurcation
structure but it becomes increasingly inaccurate at high $\alpha_{1}$
values.

\section*{Acknowledgements}

C.S. and J.J. acknowledge the financial support of the MINECO Project MOVELIGHT (PGC2018-099637-B-100
AEI/FEDER UE). S.G. acknowledges the PRIME program of the German Academic Exchange Service (DAAD) with funds from the German Federal Ministry of Education and Research (BMBF).

\appendix

\section{PDE via multiple time scales\label{app:deriv}}

We consider in this appendix the derivation of the PDE Eq.~\ref{eq:PDE_final_E}
by the use of multiple time scale analysis. Our starting point is
still Eqs.~(\ref{eq:ODE},\ref{eq:DAE}) 
\begin{eqnarray}
\varepsilon\frac{dE}{d\sigma} & = & \left[\left(1-i\alpha_{1}\right)N_{1}+\left(1-i\alpha_{2}\right)N_{2}-1\right]E+hY,\label{eq:ODE}\\
Y\left(\sigma\right) & = & \eta\left[E\left(\sigma-1\right)-Y\left(\sigma-1\right)\right],\label{eq:DAE}
\end{eqnarray}

For the sake of simplicity, we set $\alpha_{j}=0$ and assume that
$\eta$ is real. It is also convenient for our analysis to transform
the system composed of an ODE and a DAE as given by Eqs.~(\ref{eq:ODE},\ref{eq:DAE})
into a single Neutral Differential Delay Equation (NDDE) 
\begin{widetext}
\begin{equation}
\varepsilon\left[\frac{dE}{d\sigma}\left(\sigma\right)+\eta\frac{dE}{d\sigma}\left(\sigma-1\right)\right]=\left[\left(N_{1}+N_{2}\right)E\right]\left(\sigma\right)-E\left(\sigma\right)+\eta\left[\left(N_{1}+N_{2}\right)E\right]\left(\sigma-1\right)+\left(h-1\right)\eta E\left(\sigma-1\right),\label{eq:NDDE}
\end{equation}
\end{widetext}

We can define small deviations of the carriers as 
\begin{equation}
N_{j}=J_{j}+\varepsilon^{3}n_{j}
\end{equation}
with $j\in\left[1,2\right]$ and scale the field, for convenience,
as $E=\varepsilon^{\frac{3}{2}}A$. Inserting these scaling relations
in Eq.~\ref{eq:NDDE}, we find 
\begin{equation}
A\left(\sigma\right)=\eta\frac{h-1+N_{t}}{1-N_{t}}A\left(\sigma-1\right)+\frac{\varepsilon^{3}\bar{\mathcal{N}}-\varepsilon\bar{\mathcal{L}}}{1-N_{t}},
\end{equation}
with $N_{t}=J_{1}+J_{2}$ the threshold inversion given in Eq.~\ref{eq:Nth}
and the filtering $\bar{\mathcal{L}}$ and nonlinear operator $\bar{\mathcal{N}}$
defined by 
\begin{eqnarray}
\bar{\mathcal{L}} & = & \frac{d}{d\sigma}\left[A\left(\sigma\right)+\eta A\left(\sigma-1\right)\right]\\
\bar{\mathcal{N}} & = & \sum_{j=1}^{j=2}\left(n_j A\right)\left(\sigma\right)+\eta\left(n_j A\right)\left(\sigma-1\right)
\end{eqnarray}

We can now use the definition of threshold given in Eq.~\ref{eq:Nth}
to find the system upon which one can perform the multi-scale analysis
\begin{equation}
A\left(\sigma\right)=A\left(\sigma-1\right)+\frac{1+\eta}{h\eta}\left(\varepsilon^{3}\bar{\mathcal{N}}-\varepsilon\bar{\mathcal{L}}\right),\label{eq:Multi_scale_start}
\end{equation}

By inspecting Equation \ref{eq:Multi_scale_start}, one notices that
the solutions are weakly perturbed period one (P1) orbits. Because
we assume that the nonlinear term scales as $\varepsilon^{3}$, we
can safely restrict our analysis to the multi-scale analysis on the
linear part of Eq.~\ref{eq:Multi_scale_start}, that is 
\begin{equation}
A\left(\sigma\right)=A\left(\sigma-1\right)-\tilde{\varepsilon}\frac{d}{d\sigma}\left[A\left(\sigma\right)+\eta A\left(\sigma-1\right)\right],\label{eq:Multi_scale_lin}
\end{equation}
where we defined the short-hand $\tilde{\varepsilon}=\varepsilon\left(1+\eta\right)/\left(h\eta\right)$.
The linear NDDE as given by Eq.~\ref{eq:Multi_scale_lin} only depends
on one parameter, which are the cavity losses, and the smallness parameter.
It is also an excellent toy model to test various multi-scale schemes.
Notice that such a linear NDDE can be solved in the Fourier domain
directly. We, however, solve the dynamics by defining a multi-scale
expansion. The fast time is $\sigma_{0}=\sigma/T$ with $T$ the natural
period of the solution. The period $T$ shall be close to unity and
its deviation leads to the slow drift in the PDE representation. We
set $T=1+\tilde{\varepsilon}a$ where $a$ can be chosen to cancel
resonant terms in the multi-scale expansion at first order. This approach
avoids introducing altogether the intermediate time scale $\sigma_{1}=\varepsilon\sigma$.
However, one can also find analytically the period of the solution
using the functional mapping method which yields $a=1+\eta$. Finally,
we define the slow times $\sigma_{2}=\tilde{\varepsilon}^{2}\sigma$
and $\sigma_{3}=\tilde{\varepsilon}^{3}\sigma$ in order to take into
account the effect of nonlinearity, diffusion and third order dispersion.
The chain rule yields 
\begin{equation}
\frac{d}{d\sigma}\rightarrow\frac{1}{T}\frac{\partial}{\partial\sigma_{0}}+\tilde{\varepsilon}^{2}\frac{\partial}{\partial\sigma_{2}}+\tilde{\varepsilon}^{3}\frac{\partial}{\partial\sigma_{3}}
\end{equation}
while for the conjugated Fourier variables, we have 
\begin{equation}
\omega_{\sigma}\rightarrow\frac{1}{T}\omega_{0}+\tilde{\varepsilon}^{2}\omega_{2}+\tilde{\varepsilon}^{3}\omega_{3}
\end{equation}

We expand the solution as 
\begin{equation}
A\left(\sigma_{0},\sigma_{2},\sigma_{3}\right)=\sum_{j=0}^{\infty}\varepsilon^{j}A_{j}\left(\sigma_{0},\sigma_{2},\sigma_{3}\right)
\end{equation}

The various solvability conditions lead to several conditions of the
form $\bar{L}\tilde{A}\left(\omega_{0},\omega_{2},\omega_{3}\right)=0$
where $\tilde{A}\left(\omega_{0},\omega_{2},\omega_{3}\right)$ is
the (triple) Fourier transform of $A\left(\sigma_{0},\sigma_{2},\sigma_{3}\right)$.
The expression of $\bar{L}$ is 
\begin{equation}
\bar{L}\left(\omega_{\sigma}\right)=\exp\left(i\omega_{\sigma}\right)-1+i\tilde{\varepsilon}\omega_{\sigma}\left[1+\eta\exp\left(i\omega_{\sigma}\right)\right]
\end{equation}
The zeroth order operator reads naturally 
\begin{equation}
\bar{L}_{0}=e^{i\omega_{0}}-1
\end{equation}
hence showing that P1 solutions on the fast scale $\sigma_{0}$, i.e.
$A_{0}\left(\sigma_{0}-1,\sigma_{2},\sigma_{3}\right)=A_{0}\left(\sigma_{0},\sigma_{2},\sigma_{3}\right)$,
belong to the kernel of $\bar{L}_{0}$. The first order solvability
is trivially solved since $\bar{L}_{1}=0$ due to our adequate choice
of the period $T$. The other operators found at second and third
order are more complex, yet they simplify when acting on P1 solutions, 
\begin{eqnarray}
\bar{L}_{2} & = & \left(\eta^{2}-1\right)\omega_{0}^{2}+2i\omega_{2},\\
\bar{L}_{3} & = & -i\frac{\eta^{3}+1}{3}\omega_{0}^{3}-\frac{\eta-1}{2}\left(\eta+1\right)^{2}\omega_{0}^{2}+i\omega_{3}.
\end{eqnarray}

One can then build the conjugated variable to the slow time $\omega_{\xi}$
as 
\[
\omega_{\xi}=\left(\frac{1}{T}-1\right)\omega_{0}+\omega_{2}\tilde{\varepsilon}^{2}+\omega_{3}\tilde{\varepsilon}^{3}
\]
We note that adding the $-\omega_{0}$ corresponds to the stroboscopic
effect that transforms a T-periodic solution into a slowly drifting
steady state from one round-trip to the other. Upon simplification,
we find 
\[
-i\omega_{\xi}=i\left(1+\eta\right)\tilde{\omega}+\frac{\eta^{2}-1}{2}\tilde{\omega}^{2}-i\frac{\eta^{3}+1}{3}\tilde{\omega}^{3}+\mathcal{O}\left(\varepsilon\right)
\]
where we defined $\tilde{\varepsilon}\omega_{0}=\tilde{\omega}$.
Using the definition of $\tilde{\varepsilon}$ and going back to the
original variable, i.e. $\varepsilon\omega_{0}\rightarrow\omega$,
gives the following expression for the slow evolution 
\begin{equation}
\omega_{\xi}=d_{1}\omega-id_{2}\omega^{2}-d_{3}\omega^{3}
\end{equation}
with the coefficients $d_{j}$ defined in Eq.~\ref{eq:PDE_final_E}.
Finally, using that $-i\omega\rightarrow\partial_{z}$ we get 
\begin{equation}
\partial_{\xi}=d_{1}\partial_{z}+d_{2}\partial_{z}^{2}+d_{3}\partial_{z}^{3}
\end{equation}

As we assumed that the nonlinear term scales as $\varepsilon^{3}$,
it can simply be added to the highest order solvability condition
found for the operator $\bar{L}_{3}$. By using that the carriers
are also P1 solutions on the fast time scale $\left(\sigma_{0}\right)$
we get that $\left(n_{j}A\right)\left(\sigma_{0}-1,\sigma_{2},\sigma_{3}\right)=\left(n_{j}A\right)\left(\sigma_{0},\sigma_{2},\sigma_{3}\right)$
with $j\in\left[1,2\right]$ which allows finding the PDE in the main
manuscript.


\begin{thebibliography}{10}

\bibitem{MC-APL-65}
Hans~W. Mocker and R.~J. Collins.
\newblock Mode competition and self-locking effects in a {$Q$}-switched ruby
  laser.
\newblock {\em Applied Physics Letters}, 7(10):270--273, 1965.

\bibitem{lorenser04}
D.~Lorenser, H.~J. Unold, D.~J. H.~C. Maas, A.~Aschwanden, R.~Grange,
  R.~Paschotta, D.~Ebling, E.~Gini, and U.~Keller.
\newblock Towards wafer-scale integration of high repetition rate passively
  mode-locked surface-emitting semiconductor lasers.
\newblock {\em Appl. Phys. B}, 79:927--932, 2004.

\bibitem{keller96}
U.~Keller, K.~J. Weingarten, F.~X. K{\"a}rtner, D.~Kopf, B.~Braun, I.~D. Jung,
  R.~Fluck, C.~H{\"o}nninger, N.~Matuschek, and J.~Aus~der Au.
\newblock Semiconductor saturable absorber mirrors {(SESAM's)} for femtosecond
  to nanosecond pulse generation in solid-state lasers.
\newblock {\em Selected Topics in Quantum Electronics, IEEE Journal of},
  2:435--453, 1996.

\bibitem{GP-PRL-02}
A.~Gordon and B.~Fischer.
\newblock Phase transition theory of many-mode ordering and pulse formation in
  lasers.
\newblock {\em Physical Review Letters}, 89:103901--3, 2002.

\bibitem{WRG-PRL-05}
R.~Weill, A.~Rosen, A.~Gordon, O.~Gat, and B.~Fischer.
\newblock Critical behavior of light in mode-locked lasers.
\newblock {\em Physical Review Letters}, 95:013903, 2005.

\bibitem{Innoptics}
Arnaud Garnache, Vincent Lecocq, Laurence Ferrières, Attia Benselama, Mikhaël
  Myara, Laurent Cerutti, Isabelle Sagnes, and Stéphane Denet.
\newblock Industrial integration of high coherence tunable vecsel in the nir
  and mir.
\newblock {\em Proc.SPIE}, 8966:8966 -- 8966 -- 10, 2014.

\bibitem{DMB-JSTQE-13}
M.~Devautour, A.~Michon, G.~Beaudoin, I.~Sagnes, L.~Cerutti, and A.~Garnache.
\newblock Thermal management for high-power single-frequency tunable
  diode-pumped vecsel emitting in the near- and mid-ir.
\newblock {\em IEEE Journal of Selected Topics in Quantum Electronics},
  19(4):1701108--1701108, July 2013.

\bibitem{lidar}
N.~Takeuchi, N.~Sugimoto, H.~Baba, and K.~Sakurai.
\newblock Random modulation {cw lidar}.
\newblock {\em Appl. Opt.}, 22(9):1382--1386, May 1983.

\bibitem{lidar2}
Nobuo Takeuchi, Hiroshi Baba, Katsumi Sakurai, and Toshiyuki Ueno.
\newblock Diode-laser random-modulation cw lidar.
\newblock {\em Applied Optics}, 25:63--67, 1985.

\bibitem{haus00rev}
H.~A. Haus.
\newblock Mode-locking of lasers.
\newblock {\em IEEE J. Selected Topics Quantum Electron.}, 6:1173--1185, 2000.

\bibitem{AJ-BOOK-17}
E.~Avrutin and J.~Javaloyes.
\newblock {\em Mode-Locked Semiconductor Lasers, Book Chapter In: Handbook of
  Optoelectronic Device Modeling and Simulation}.
\newblock CRC press, Taylor and Francis, United Kingdom, 2017.

\bibitem{LMW-SCI-17}
S.~M. Link, D.~J. H.~C. Maas, D.~Waldburger, and U.~Keller.
\newblock Dual-comb spectroscopy of water vapor with a free-running
  semiconductor disk laser.
\newblock {\em Science}, 356(6343):1164--1168, 2017.

\bibitem{UHH-NAT-02}
Th~Udem, R.~Holzwarth, and T.~W. H{\"a}nsch.
\newblock Optical frequency metrology.
\newblock {\em Nature}, 416(6877):233--237, 2002.

\bibitem{hoogland00}
S.~Hoogland, S.~Dhanjal, A.~C. Tropper, J.~S. Roberts, R.~H{\"a}ring,
  R.~Paschotta, F.~Morier-Genoud, and U.~Keller.
\newblock Passively mode-locked diode-pumped surface-emitting semiconductor
  lasers.
\newblock {\em IEEE Photonics Technology Letters}, 12:1135--1137, 2000.

\bibitem{haring01}
R.~Haring, R.~Paschotta, E.~Gini, F.~Morier-Genoud, D.~Martin, H.~Melchior, and
  U.~Keller.
\newblock Picosecond surface-emitting semiconductor laser with {$> 200$} mw
  average power.
\newblock {\em Electronics Letters}, 37(12):766--767, Jun 2001.

\bibitem{haring02}
R.~H{\"a}ring, R.~Paschotta, A.~Aschwanden, E.~Gini, F.~Morier-Genoud, and
  U.~Keller.
\newblock High-power passively mode-locked semiconductor lasers.
\newblock {\em Quantum Electronics, IEEE Journal of}, 38:1268--1275, 2002.

\bibitem{RRH-OL-08}
B.~Rudin, A.~Rutz, M.~Hoffmann, D.~J. H.~C. Maas, A.-R. Bellancourt, E.~Gini,
  T.~S\"{u}dmeyer, and U.~Keller.
\newblock Highly efficient optically pumped vertical-emitting semiconductor
  laser with more than 20 w average output power in a fundamental transverse
  mode.
\newblock {\em Opt. Lett.}, 33(22):2719--2721, Nov 2008.

\bibitem{WLM-OptA-16}
Dominik Waldburger, Sandro~M. Link, Mario Mangold, Cesare G.~E. Alfieri, Emilio
  Gini, Matthias Golling, Bauke~W. Tilma, and Ursula Keller.
\newblock High-power 100 fs semiconductor disk lasers.
\newblock {\em Optica}, 3(8):844--852, Aug 2016.

\bibitem{MBR-APB-07}
D.J.H.C. Maas, A.-R. Bellancourt, B.~Rudin, M.~Golling, H.J. Unold,
  T.~S{\"u}dmeyer, and U.~Keller.
\newblock Vertical integration of ultrafast semiconductor lasers.
\newblock {\em Applied Physics B}, 88(4):493--497, Sep 2007.

\bibitem{RWM-OE-10}
B.~Rudin, V.~J. Wittwer, D.~J. H.~C. Maas, M.~Hoffmann, O.~D. Sieber,
  Y.~Barbarin, M.~Golling, T.~S\"{u}dmeyer, and U.~Keller.
\newblock High-power mixsel: an integrated ultrafast semiconductor laser with
  6.4 w average power.
\newblock {\em Opt. Express}, 18(26):27582--27588, Dec 2010.

\bibitem{CSV-OL-18}
P.~Camelin, C.~Schelte, A.~Verschelde, A.~Garnache, G.~Beaudoin, I.~Sagnes,
  G.~Huyet, J.~Javaloyes, S.~V. Gurevich, and M.~Giudici.
\newblock Temporal localized structures in mode-locked vertical external-cavity
  surface-emitting lasers.
\newblock {\em Opt. Lett.}, 43(21):5367--5370, Nov 2018.

\bibitem{SCM_PRL_19}
C.~Schelte, P.~Camelin, M.~Marconi, A.~Garnache, G.~Huyet, G.~Beaudoin,
  I.~Sagnes, M.~Giudici, J.~Javaloyes, and S.~V. Gurevich.
\newblock Third order dispersion in time-delayed systems.
\newblock {\em Phys. Rev. Lett.}, 123:043902, Jul 2019.

\bibitem{JNS-PRE-16}
Lina Jaurigue, Oleg Nikiforov, Eckehard Sch\"oll, Stefan Breuer, and Kathy
  L\"udge.
\newblock Dynamics of a passively mode-locked semiconductor laser subject to
  dual-cavity optical feedback.
\newblock {\em Phys. Rev. E}, 93:022205, Feb 2016.

\bibitem{JKL-CHA-17}
Lina Jaurigue, Bernd Krauskopf, and Kathy Lüdge.
\newblock Multipulse dynamics of a passively mode-locked semiconductor laser
  with delayed optical feedback.
\newblock {\em Chaos: An Interdisciplinary Journal of Nonlinear Science},
  27(11):114301, 2017.

\bibitem{AHP-JOSAB-16}
R.~M. Arkhipov, T.~Habruseva, A.~Pimenov, M.~Radziunas, S.~P. Hegarty,
  G.~Huyet, and A.~G. Vladimirov.
\newblock Semiconductor mode-locked lasers with coherent dual-mode optical
  injection: simulations, analysis, and experiment.
\newblock {\em J. Opt. Soc. Am. B}, 33(3):351--359, Mar 2016.

\bibitem{SCB-PRL-19}
Yifan~Sun Sun, Sylvain Combri\'e, Fabien Bretenaker, and Alfredo De~Rossi.
\newblock Mode locking of the hermite-gaussian modes of a nanolaser.
\newblock {\em Phys. Rev. Lett.}, in print, 2019.

\bibitem{RBM-JQE-11}
M.~Rossetti, P.~Bardella, and I.~Montrosset.
\newblock Modeling passive mode-locking in quantum dot lasers: A comparison
  between a finite-difference traveling-wave model and a delayed differential
  equation approach.
\newblock {\em Quantum Electronics, IEEE Journal of}, 47(5):569 --576, may
  2011.

\bibitem{BSR-BOOK-14}
Stefan Breuer, Dimitris Syvridis, and Edik~U. Rafailov.
\newblock {\em Ultra-Short-Pulse QD Edge-Emitting Lasers}, pages 43--94.
\newblock Wiley-VCH Verlag GmbH \& Co. KGaA, 2014.

\bibitem{AWN-PRAp-18}
C.~G.~E. Alfieri, D.~Waldburger, J.~N\"urnberg, M.~Golling, L.~Jaurigue,
  K.~L\"udge, and U.~Keller.
\newblock Mode-locking instabilities for high-gain semiconductor disk lasers
  based on active submonolayer quantum dots.
\newblock {\em Phys. Rev. Applied}, 10:044015, Oct 2018.

\bibitem{MLA-OE-18}
T.~Malica, J.~Lin, T.~Ackemann, D.~J. Little, J.~P. Toomey, D.~Pab{\oe}uf,
  W.~Lubeigt, N.~Hempler, G.~Malcolm, G.~T. Maker, and D.~M. Kane.
\newblock Mapping the dynamical regimes of a sesam mode-locked vecsel with a
  long cavity using time series analysis.
\newblock {\em Opt. Express}, 26(13):16624--16638, Jun 2018.

\bibitem{HML-PRAp-19}
Jan Hausen, Stefan Meinecke, Benjamin Lingnau, and Kathy L\"udge.
\newblock Pulse cluster dynamics in passively mode-locked semiconductor
  vertical-external-cavity surface-emitting lasers.
\newblock {\em Phys. Rev. Applied}, 11:044055, Apr 2019.

\bibitem{MJB-PRL-14}
M.~Marconi, J.~Javaloyes, S.~Balle, and M.~Giudici.
\newblock How lasing localized structures evolve out of passive mode locking.
\newblock {\em Phys. Rev. Lett.}, 112:223901, Jun 2014.

\bibitem{CJM-PRA-16}
P.~Camelin, J.~Javaloyes, M.~Marconi, and M.~Giudici.
\newblock Electrical addressing and temporal tweezing of localized pulses in
  passively-mode-locked semiconductor lasers.
\newblock {\em Phys. Rev. A}, 94:063854, Dec 2016.

\bibitem{JCM-PRL-16}
J.~Javaloyes, P.~Camelin, M.~Marconi, and M.~Giudici.
\newblock Dynamics of localized structures in systems with broken parity
  symmetry.
\newblock {\em Phys. Rev. Lett.}, 116:133901, Mar 2016.

\bibitem{YRS-PRL-19}
Serhiy Yanchuk, Stefan Ruschel, Jan Sieber, and Matthias Wolfrum.
\newblock Temporal dissipative solitons in time-delay feedback systems.
\newblock {\em Phys. Rev. Lett.}, 123:053901, Jul 2019.

\bibitem{SPH-NAP-12}
Albert Schliesser, Nathalie Picqu{\'e}, and Theodor~W. H{\"a}nsch.
\newblock Mid-infrared frequency combs.
\newblock {\em Nature Photonics}, 6(7):440--449, 2012.

\bibitem{K-NAT-03}
Ursula Keller.
\newblock Recent developments in compact ultrafast lasers.
\newblock {\em Nature}, 424(6950):831--838, 2003.

\bibitem{J-PRL-16}
J.~Javaloyes.
\newblock Cavity light bullets in passively mode-locked semiconductor lasers.
\newblock {\em Phys. Rev. Lett.}, 116:043901, Jan 2016.

\bibitem{GJ-PRA-17}
S.~V. Gurevich and J.~Javaloyes.
\newblock Spatial instabilities of light bullets in passively-mode-locked
  lasers.
\newblock {\em Phys. Rev. A}, 96:023821, Aug 2017.

\bibitem{SJG-OL-18}
C.~Schelte, J.~Javaloyes, and S.~V. Gurevich.
\newblock Functional mapping for passively mode-locked semiconductor lasers.
\newblock {\em Opt. Lett.}, 43(11):2535--2538, Jun 2018.

\bibitem{GT-CRA-64}
F.~Gires and P.~Tournois.
\newblock Interferometre utilisable pour la compression d'impulsions lumineuses
  modulees en frequence.
\newblock {\em C. R. Acad. Sci. Paris}, (258):6112--6115, 1964.

\bibitem{MB-JQE-05}
J.~Mulet and S.~Balle.
\newblock Mode locking dynamics in electrically-driven vertical-external-cavity
  surface-emitting lasers.
\newblock {\em Quantum Electronics, IEEE Journal of}, 41(9):1148--1156, 2005.

\bibitem{NF-BOOK}
Ali~H Nayfeh.
\newblock {\em The Method of Normal Forms}.
\newblock John Wiley \& Sons, Hoboken, NJ, 2011.

\bibitem{uecker2014}
Hannes Uecker, Daniel Wetzel, and Jens D.~M. Rademacher.
\newblock pde2path - a matlab package for continuation and bifurcation in 2d
  elliptic systems.
\newblock {\em Numerical Mathematics: Theory, Methods and Applications},
  7(1):58--106, 002 2014.

\bibitem{VT-PRA-05}
A.~G. Vladimirov and D.~Turaev.
\newblock Model for passive mode locking in semiconductor lasers.
\newblock {\em Phys. Rev. A}, 72:033808, Sep 2005.

\bibitem{GP-PRL-96}
G.~Giacomelli and A.~Politi.
\newblock Relationship between delayed and spatially extended dynamical
  systems.
\newblock {\em Phys. Rev. Lett.}, 76:2686--2689, Apr 1996.

\bibitem{K-CMMP-98}
S.A. Kashchenko.
\newblock {The Ginzburg-Landau equation as a normal form for a second-order
  difference-differential equation with a large delay.}
\newblock {\em {Comput. Math. Math. Phys.}}, 38(3):1, 1998.

\bibitem{GMZ-PRE-13}
Giovanni Giacomelli, Francesco Marino, Michael~A. Zaks, and Serhiy Yanchuk.
\newblock Nucleation in bistable dynamical systems with long delay.
\newblock {\em Phys. Rev. E}, 88:062920, Dec 2013.

\bibitem{LPM-PRL-13}
Laurent Larger, Bogdan Penkovsky, and Yuri Maistrenko.
\newblock Virtual chimera states for delayed-feedback systems.
\newblock {\em Phys. Rev. Lett.}, 111:054103, Aug 2013.

\bibitem{MGB-PRL-14}
Francesco Marino, Giovanni Giacomelli, and Stephane Barland.
\newblock Front pinning and localized states analogues in long-delayed bistable
  systems.
\newblock {\em Phys. Rev. Lett.}, 112:103901, Mar 2014.

\bibitem{YG-PRL-14}
Serhiy Yanchuk and Giovanni Giacomelli.
\newblock Pattern formation in systems with multiple delayed feedbacks.
\newblock {\em Phys. Rev. Lett.}, 112:174103, May 2014.

\bibitem{EJW-CHA-17}
Thomas Erneux, Julien Javaloyes, Matthias Wolfrum, and Serhiy Yanchuk.
\newblock Introduction to focus issue: Time-delay dynamics.
\newblock {\em Chaos: An Interdisciplinary Journal of Nonlinear Science},
  27(11):114201, 2017.

\bibitem{YG-JPA-17}
Serhiy Yanchuk and Giovanni Giacomelli.
\newblock Spatio-temporal phenomena in complex systems with time delays.
\newblock {\em Journal of Physics A: Mathematical and Theoretical},
  50(10):103001, 2017.

\bibitem{LK-JQE-80}
R.~Lang and K.~Kobayashi.
\newblock External optical feedback effects on semiconductor injection laser
  properties.
\newblock {\em Quantum Electronics, IEEE Journal of}, 16(3):347 -- 355, mar
  1980.

\bibitem{DDEBT}
K.~Engelborghs, T.~Luzyanina, and D.~Roose.
\newblock Numerical bifurcation analysis of delay differential equations using
  dde-biftool.
\newblock {\em ACM Trans. Math. Softw.}, 28(1):1--21, March 2002.

\bibitem{SJG-PRA-18}
C.~Schelte, J.~Javaloyes, and S.~V. Gurevich.
\newblock Dynamics of temporally localized states in passively mode-locked
  semiconductor lasers.
\newblock {\em Phys. Rev. A}, 97:053820, May 2018.

\bibitem{KNE-PD-06}
Theodore Kolokolnikov, Michel Nizette, Thomas Erneux, Nicolas Joly, and Serge
  Bielawski.
\newblock The {$Q$}-switching instability in passively mode-locked lasers.
\newblock {\em Physica D: Nonlinear Phenomena}, 219(1):13 -- 21, 2006.

\end{thebibliography}

\end{document}